\documentclass[
amsmath,
prd,nofootinbib,floatfix,12pt,
]{revtex4}

\usepackage{epsf}

\newcommand\beq{\begin{eqnarray}}
\newcommand\eeq{\end{eqnarray}}

\def\lsim{\mathrel{\rlap{\lower4pt\hbox{$\sim$}}
    \raise1pt\hbox{$<$}}}                
\def\gsim{\mathrel{\rlap{\lower4pt\hbox{$\sim$}}
    \raise1pt\hbox{$>$}}}            

\allowdisplaybreaks
\usepackage{graphicx}
\usepackage{setspace}
\usepackage{dcolumn}
\usepackage{bm}

\usepackage[dvips]{color}
\definecolor{Red}{cmyk}{0,1,1,0}
\definecolor{BrickRed}{cmyk}{0,0.89,0.94,0.28}
\definecolor{Blue}{cmyk}{1,1,0,0}
\definecolor{Green}{cmyk}{1,0,1,0}

\newcommand\MSbar{\overline{{\rm MS}}}
\newcommand\lnbar{\overline{\ln}}

\begin{document}

\title{\large \baselineskip=20pt 
Matching relations for decoupling in the Standard Model\\ at two loops and beyond}

\author{Stephen P.~Martin}
\affiliation{\it Department of Physics, Northern Illinois University, DeKalb IL 60115}

\begin{abstract}\normalsize \baselineskip=17pt
I discuss the matching relations for the running renormalizable parameters when the heavy particles 
(top quark, Higgs scalar, $Z$ and $W$ vector bosons) are simultaneously decoupled
from the Standard Model. The complete two-loop order matching for the
electromagnetic coupling and all light fermion masses are obtained,
augmenting existing results at 4-loop order in pure
QCD and complete two-loop order for the strong coupling.
I also review the further sequential decouplings of 
the lighter fermions (bottom quark, tau lepton, and charm quark)
from the low-energy effective theory.
\end{abstract}

\maketitle

\vspace{-0.3in}

\tableofcontents
\baselineskip=16.4pt
\newpage

\section{Introduction \label{sec:intro}}
\setcounter{equation}{0}
\setcounter{figure}{0}
\setcounter{table}{0}
\setcounter{footnote}{1}

The discovery of the Higgs scalar boson at the Large Hadron Collider has 
put the Standard Model of particle physics on a firm footing. At the same time, 
searches for physics beyond the Standard Model have not produced 
confirmed hints of any more fundamental structure. 
It therefore seems worthwhile to consider the Standard Model as 
quite possibly valid and complete 
up to well above the TeV energy scale,
and to study its precise parameters and predictions, 
assuming that the next layer of fundamental 
new physics particles is heavy enough to be irrelevant at energy scales 
now within direct reach at colliders.

The Standard Model has within it an interesting hierarchy, with four fundamental particles
(the top quark, the Higgs scalar, and the $Z$ and $W$ vector
bosons) having masses
within a factor of 2.2 each other, and heavier than all others 
by well over an order of magnitude. This makes it sensible to consider a low-energy effective
theory consisting of the $b,c,s,u,d$ quarks, the $\tau,\mu,e$ leptons, 
and their neutrinos, with renormalizable interactions coming from the unbroken
$SU(3)_c \times U(1)_{\rm EM}$ gauge group, 
and non-renormalizable four-fermion couplings to describe the weak interactions.
This low-energy effective field theory can be matched onto the full 
$SU(3)_c \times SU(2)_L \times U(1)_Y$ high-energy theory with no particles
decoupled, by considering common physical observables calculated in each theory
in terms of parameters defined in the $\MSbar$ renormalization scheme 
\cite{Bardeen:1978yd,Braaten:1981dv} 
based on dimensional regularization 
\cite{Bollini:1972bi,Ashmore:1972uj,Cicuta:1972jf,tHooft:1972tcz,tHooft:1973mfk}.

In this paper, I will consider the decoupling relations that govern the matching
at an arbitrary $\MSbar$ renormalization scale, denoted $Q$.
Specifically, the pertinent running $\MSbar$ 
parameters of the full Standard Model will be called
\beq
g_3,\> g,\> g',\> 
y_t,\> y_b,\> y_c,\> y_s,\> y_u,\> y_d,\> y_\tau,\> y_\mu,\> y_e,\> 
\lambda,\> v.
\label{eq:highenergyparams}
\eeq
Here, $g_3$, $g$, and $g'$ are the gauge couplings,
the $y_f$ are the Yukawa couplings, $\lambda$ is the Higgs 
self-interaction coupling, and $v$ is the Higgs vacuum expectation value
(VEV), defined
in this paper as the minimum of the effective potential in Landau gauge.
This definition implies that scalar tadpole sub-graphs vanish identically when 
summed to all orders in perturbation theory (including the tree-level tadpole), and so can be 
omitted from all Feynman diagrams.\footnote{The Landau gauge
Standard Model effective potential and its minimization condition
are presently known to full 2-loop \cite{Ford:1992pn,Martin:2001vx} and
3-loop \cite{Martin:2013gka,Martin:2017lqn} orders,
and the 4-loop part only at leading order in QCD \cite{Martin:2015eia}.
These results make use of
Goldstone boson resummation \cite{Martin:2014bca,Elias-Miro:2014pca},
and employ 3-loop vacuum integral basis functions defined and evaluated by 
\cite{Martin:2016bgz}; for an alternative evaluation method
see \cite{Freitas:2016zmy}.
In particular, refs.~\cite{Martin:2017lqn,Martin:2015eia} 
provide the formulas relating the VEV $v$ used here to 
the tree-level VEV
$v_{\rm tree} = \sqrt{-m_H^2/\lambda}$ used in many other works, 
which therefore must \cite{Fleischer:1980ub} include tadpole graphs.
Outside of Landau gauge, the effective potential 
is much more complicated at 2-loop order \cite{Martin:2018emo}, 
and not known at 3-loop order.}
The very small effects of 
Cabibbo-Kobayashi-Maskawa mixing and neutrino masses are neglected. 
The running $\MSbar$ squared masses of the Standard Model 
states are then denoted:
\beq
Z &=& (g^2 + g^{\prime 2}) v^2/4,
\label{eq:defZ}
\\
W &=& g^2 v^2/4,
\\
h &=& 2 \lambda v^2,
\\
t &=& y_t^2 v^2/2,
\label{eq:deft}
\\
b &=& y_b^2 v^2/2,\qquad {\rm etc.}
\eeq
Due to the choice of the definition of the VEV $v$, these quantities are 
specific to Landau gauge. 
As a matter of preference, I find the convenience 
(and increased accuracy) of not having tadpole graphs 
(with their associated $1/\lambda$ factors in perturbation theory,
coming from zero-momentum Higgs propagators)
to be well worth the price of a Landau-gauge-specific VEV and running masses,
especially since these are not renormalization group scale-invariant 
observables anyway.
The high-energy non-decoupled electromagnetic coupling is defined by
\beq
e \equiv g g^{\prime}/\sqrt{g^2 + g^{\prime 2}}.
\label{eq:defe}
\eeq 

In the low-energy $SU(3)_c \times U(1)_{\rm EM}$ effective field theory, 
the renormalizable $\MSbar$ parameters will be denoted 
in this paper as 
\beq
\alpha_S,\> \alpha,\> m_b,\> m_c,\> m_s,\> m_u,\> m_d,\> m_\tau,\> m_\mu,\> m_e.
\label{eq:lowenergyparams}
\eeq
To avoid confusion, $\alpha_S$ and $\alpha$ are only used for the
low-energy effective theory, and never for
the gauge couplings of the non-decoupled full Standard Model theory.
Conversely, the symbols 
$g_3$, $g$, $g'$, and $e$ are used exclusively to refer to quantities
in the full non-decoupled theory. Note also that $\alpha$ is 
used in this paper to refer to the $\MSbar$ quantity, not the so-called 
``on-shell" electromagnetic coupling.
All of the parameters in 
eqs.~(\ref{eq:highenergyparams})-(\ref{eq:lowenergyparams}) depend on the
$\MSbar$ renormalization scale $Q$.

There are several complementary paths that one can take to relating 
these parameters to experimental results. In one approach, one makes direct use of 
low-energy experimental observables as the basic inputs,
which then determine the parameters in eq.~(\ref{eq:lowenergyparams}), 
and then infer the full Standard Model parameters in eq.~(\ref{eq:highenergyparams}) from them. 
In this paper, I will instead take the basic input parameters to be
those of eq.~(\ref{eq:highenergyparams}); then the low-energy observable data
can be derived and used as the subjects of global fits. 
The purpose of this paper is limited to finding the matching relations that give
the parameters of eq.~(\ref{eq:lowenergyparams}) as functions of those in
eq.~(\ref{eq:highenergyparams}). This will be done treating the matching scale
$Q$ as arbitrary, with the assumption that, typically, 
it should be chosen not much smaller than
the $W$-boson mass and not much larger than the top-quark mass, in order to
avoid unnecessary large logarithms. Note that $\ln(M_t/M_W) = 0.77$, so
that any choice of $M_W \lsim Q \lsim M_t$ for the matching scale 
should be fine. (It is not necessary that each particle is 
automatically decoupled at the scale $Q$ equal to its mass, 
which is ambiguous in any case.)

Some observables, notably the pole masses of the top, Higgs, $Z$, and $W$,
are only accessible in the high energy theory. 
The Higgs boson mass has been connected to the self-coupling $\lambda$
including 2-loop QCD corrections \cite{Bezrukov:2012sa} 
and at full 2-loop order 
in terms of interpolating formulas \cite{Degrassi:2012ry,Buttazzo:2013uya}.
Analytical results and computer code for the Higgs mass at complete 
2-loop order have been presented in the tadpole-free scheme 
consistent with the present paper in
ref.~\cite{Martin:2014cxa}, which also includes leading 3-loop corrections, 
and in the scheme with a tree-level VEV and tadpoles in 
refs.~\cite{Kniehl:2015nwa,Kniehl:2016enc}.
Multi-loop corrections to the
$W$ and $Z$ boson masses, their ratio (the $\rho$ parameter), 
and their relationships with other observables have been discussed in 
\cite{Sirlin:1980nh,Marciano:1980pb,Marciano:1983wwa,Sirlin:1983ys,Djouadi:1987gn,Djouadi:1987di,Consoli:1989fg,Kniehl:1989yc,Halzen:1990je,Djouadi:1993ss,Avdeev:1994db,Chetyrkin:1995ix,Chetyrkin:1995js,Degrassi:1996mg,Degrassi:1996ps,Degrassi:1997iy,Passera:1998uj,Freitas:2000gg,Freitas:2002ja,Awramik:2002wn,Onishchenko:2002ve,Awramik:2002vu,Faisst:2003px,Awramik:2003ee,Awramik:2003rn,Schroder:2005db, Chetyrkin:2006bj,Boughezal:2006xk,Jegerlehner:2001fb,Jegerlehner:2002er,Jegerlehner:2002em,Degrassi:2014sxa,Martin:2015lxa,Martin:2015rea},\cite{Kniehl:2015nwa,Kniehl:2016enc}. In particular, 
refs.~\cite{Martin:2015lxa,Martin:2015rea}
provide the complete 2-loop analytic
results for the $W$ and $Z$ pole masses, respectively, in the 
tadpole-free $\MSbar$ scheme consistent with the conventions and notations 
of the present paper.
For the top-quark pole mass,
the pure QCD contributions are known at 
1-loop \cite{Tarrach:1980up}, 
2-loop \cite{Gray:1990yh},
3-loop \cite{Chetyrkin:1999ys,Chetyrkin:1999qi,Melnikov:2000qh},
and 4-loop \cite{Marquard:2015qpa,Marquard:2016dcn} orders;
these results also apply to the light quark pole masses in the decoupled theory.
Contributions and uncertainty estimates from higher orders in QCD are discussed in 
\cite{Beneke:1994qe,Ball:1995ni,Kataev:2015gvt,Beneke:2016cbu,Hoang:2017btd}.
The non-QCD 1-loop corrections to fermion pole masses
were given in \cite{Bohm:1986rj,Hempfling:1994ar}.
Mixed 2-loop QCD corrections to the top-quark pole mass were obtained in 
refs.~\cite{Jegerlehner:2003py,Jegerlehner:2003sp,Faisst:2004gn,Eiras:2005yt,Jegerlehner:2012kn}, 
and the 2-loop electroweak corrections in the ``gaugeless" limit (where $W,Z$ masses are neglected 
compared to the top-quark mass) are given in 
refs.~\cite{Martin:2005ch,Kniehl:2014yia}.
The full 2-loop top-quark pole mass corrections have been given in
the tree-level-VEV scheme in \cite{Kniehl:2015nwa}, and in the tadpole-free
scheme used in the present paper in ref.~\cite{Martin:2016xsp}.

For computations at characteristic energies much lower or much higher 
than the matching scale, 
one should use the renormalization group equations to run the
$\MSbar$ parameters to an appropriate comparable $Q$, 
thus resumming the potentially large logarithms that would otherwise occur. 
For the full Standard Model, 
the beta functions are presently known
at full 2-loop \cite{MVI,MVII,Jack:1984vj,MVIII,Luo:2002ey} and 3-loop
\cite{Mihaila:2012fm,Chetyrkin:2012rz,Bednyakov:2012rb,Bednyakov:2012en,Chetyrkin:2013wya,Bednyakov:2013eba,Bednyakov:2013cpa,Bednyakov:2014pia}
orders. The beta function for the Higgs self-coupling is also known at 4 loops
in the leading order in QCD \cite{Martin:2015eia,Chetyrkin:2016ruf}.
For the strong gauge coupling, the pure QCD contributions to the beta function
are known at 4-loop \cite{vanRitbergen:1997va,Czakon:2004bu} and 5-loop
\cite{Baikov:2016tgj,Herzog:2017ohr} orders, and the
QCD contributions to the beta functions of the quark Yukawa couplings
(or equivalently, the running quark masses) are likewise known
at 3-loop \cite{Tarasov},
4-loop \cite{Chetyrkin:1997dh,Vermaseren:1997fq}, and 5-loop
\cite{Baikov:2014qja} orders.
These QCD results also apply to the $\alpha_S$ and quark masses of the
low-energy effective theory, by changing the variable number of active quarks.
 
There are also already extensive multi-loop 
results on the decoupling matching relations involving the strong interactions. 
The 1-loop and 2-loop decoupling of the QCD coupling
at quark thresholds were discussed long ago in refs.~\cite{Weinberg:1980wa}, 
and \cite{Bernreuther:1981sg,Larin:1994va}, respectively. The pure QCD 3-loop 
and 4-loop threshold corrections for $\alpha_S$
were obtained in refs.~\cite{Chetyrkin:1997un,Grozin:2011nk}
and \cite{Schroder:2005hy,Chetyrkin:2005ia}, respectively. 
The complete 2-loop 
threshold corrections for $\alpha_S$ 
including electroweak and top-quark Yukawa effects
were given in ref.~\cite{Bednyakov:2014fua},
and have been checked as part of the present work.
For the pure 
QCD contributions to quark mass threshold relations, the 3-loop results
were obtained in ref.~\cite{Chetyrkin:1997un,Grozin:2011nk}, 
and the 4-loop results in
ref.~\cite{Liu:2015fxa}. All of the pure QCD results for running and decoupling
of $\alpha_S$ and quark masses have been incorporated into 
the RunDec \cite{RunDec} computer software packages.

The electromagnetic coupling is usually related to the very precisely known 
low-energy Thomson scattering
value $\alpha_{\rm Thomson} = 1/137.0359991\ldots$ as the basic input parameter,
through radiative corrections to 
the photon self-energy function 
\cite{Sirlin:1980nh,Kniehl:1989yc,Marciano:1990dp,Degrassi:1990tu,Fanchiotti:1992tu,Eidelman:1995ny,Burkhardt:1995tt,Martin:1994we,Alemany:1997tn,Davier:1997vd,Kuhn:1998ze,Steinhauser:1998rq,Erler:1998sy,Erler:1999ug,Degrassi:2003rw,Tanabashi:2018oca},
\cite{Degrassi:2014sxa,Kniehl:2015nwa}.
The bottleneck to accuracy in running $\alpha$ to very high energies
(where it can be matched to $g, g'$) 
comes from the non-perturbative hadronic contributions,
often parameterized as $\Delta \alpha_{\rm had}^{(5)}(m_Z)$. 
For recent evaluations of this important
quantity, see refs.~\cite{Davier:2017zfy,Jegerlehner:2017zsb,Keshavarzi:2018mgv}
and references therein. In this paper, I will instead concentrate on 
the connection to the far-ultraviolet, fundamental definition of the 
Standard Model, by obtaining the complete 2-loop relationship between the
$\MSbar$ parameters $g,g',\ldots$ of the Standard Model and the
$\MSbar$ running coupling $\alpha(Q)$ in the low-energy theory when $t,h,Z,W$ are simultaneously decoupled.\footnote{Note, however, that the 
$\widehat\alpha^{(5)}(M_Z)$ quoted as the $\MSbar$ coupling 
in the Review of Particle Properties
(RPP) \cite{Tanabashi:2018oca} 
decouples the top quark but not the $W$ boson, and so is not 
the same as the $\MSbar$-scheme $\alpha(Q)$ as defined here within the
5-quark, 3-lepton, $SU(3)_c \times U(1)_{\rm EM}$ gauge theory.
In fact, strictly speaking $\widehat\alpha^{(5)}$ as defined in the 
RPP (following refs.~\cite{Marciano:1990dp,Fanchiotti:1992tu})
is not really an $\MSbar$ coupling in the usual sense, 
because once the top quark has been decoupled, $SU(2)_L$ gauge invariance is explicitly and irretrievably broken, so that the $W,Z$ bosons should also be decoupled in order to have a renormalizable effective theory.}
The relationship between $\alpha(Q)$ and the very-low energy input
$\alpha_{\rm Thomson}$ is in this paper left as a separate issue,
as addressed in 
\cite{Sirlin:1980nh,Kniehl:1989yc,Marciano:1990dp,Degrassi:1990tu,Fanchiotti:1992tu,Eidelman:1995ny,Burkhardt:1995tt,Martin:1994we,Alemany:1997tn,Davier:1997vd,Kuhn:1998ze,Steinhauser:1998rq,Erler:1998sy,Erler:1999ug,Degrassi:2003rw,Tanabashi:2018oca},
\cite{Degrassi:2014sxa,Kniehl:2015nwa}.

The other new result to be obtained below is the complete 2-loop matching for 
all of the light fermion masses listed in eq.~(\ref{eq:lowenergyparams}).
The relation between the Yukawa couplings and the pole masses of the 
lightest 5 quarks were obtained to order $\alpha_S \alpha$ in 
\cite{Kniehl:2004hfa}.
In ref.~\cite{Kniehl:2014yia}, the relationship between the 
bottom quark on-shell mass and its Yukawa coupling and running mass 
were obtained at 2-loop order in the gaugeless limit,
for both a tree-level VEV scheme and for an ``on-shell" definition of the VEV, 
$v_{\rm on-shell}^2 \equiv 1/\sqrt{2} G_F$. 
This has been extended to full 2-loop order in 
ref.\cite{Kniehl:2015nwa}, with results given in terms of numerical
linear interpolation formulas. In ref.~\cite{Bednyakov:2016onn}, 
the matching formulas for decoupling were given for the bottom quark mass, 
again using numerical interpolation formulas.
In this paper, I will give the analytic results for the matching relations 
for the bottom quark as well as all other light quark masses, using 
the tadpole-free scheme to define the Standard Model VEV 
(and thus the running masses) in the non-decoupled theory.

The method used to find each decoupling matching 
relation is to compute a gauge-invariant physical
quantity two ways, in terms of the parameters of 
the decoupled and the non-decoupled theories, and then 
require that the results agree. For the gauge couplings, 
the physical quantity is the residue of the
pole in a scattering amplitude at $p^2 = 0$, where $p^\mu$ is the 
4-momentum of the gauge boson mediating the interaction. 
In the case of the fermion masses $m_f$, the physical quantity 
is the pole mass. The methods used here for the necessary calculations
are very similar to those in 
refs.~\cite{Martin:2014cxa,Martin:2015lxa,Martin:2015rea,Martin:2016xsp,Martin:2017lqn}, 
and all notations are chosen to be 
consistent with those papers. In particular, logarithms involving
the renormalization scale will be denoted by
\beq
\lnbar(x) &\equiv& \ln(x/Q^2)
,
\label{eq:deflnbar}
\eeq
where $x = t,h,Z,W,\ldots$ are $\MSbar$ squared masses. In the following, 
only vacuum graphs
occur in the final results, so they can be written in 
terms of $\lnbar(x)$ and the 2-loop renormalized vacuum 
basis integral function \cite{Ford:1992pn,Davydychev:1992mt}.
The notation used here is, in terms of dilogarithms, for $x,y\leq z$:
\beq
I(x,y,z) &=& 
  \frac{1}{2}(x-y-z) \,\lnbar(y) \,\lnbar(z)
+ \frac{1}{2}(y-x-z) \,\lnbar(x) \,\lnbar(z)
+ \frac{1}{2}(z-x-y) \,\lnbar(x) \,\lnbar(y)
\phantom{x}
\nonumber \\ &&
+ 2 x \,\lnbar(x)
+ 2 y \,\lnbar(y)
+ 2 z \,\lnbar(z)
- \frac{5}{2} (x+y+z)
\nonumber \\ &&
+ 
r \left [
{\rm Li}_2(k_1)
+ {\rm Li}_2(k_2)
-\,\ln(k_1) \, \ln(k_2)
+ \frac{1}{2} \ln(x/z)\, \ln(y/z)
- \zeta_2 \right ]
,
\eeq
with $r = \sqrt{\lambda(x,y,z)}$ and $k_1 = (z+x-y-r)/2z$ and
$k_2 = (z+y-x-r)/2z$, where
\beq
\lambda(x,y,z) &\equiv& x^2 + y^2 + z^2 - 2 x y - 2 x z - 2 y z .
\eeq
The function $I(x,y,z)$ implicitly depends on $Q$ through the $\lnbar$
functions, and it 
is symmetric under interchange of any of its arguments $x,y,z$.
Some useful special cases are:
\beq
I(0,0,x) &=& -\frac{1}{2} x \,\lnbar^2(x) + 2 x \,\lnbar(x) - \frac{5}{2}x 
- \zeta_2 x,
\\
I(0,x,x) &=& -x \,\lnbar^2(x) + 4 x \,\lnbar(x) - 5 x,
\\
I(0,x,y) &=& (y-x) \left [{\rm Li}_2(1-y/x) + \frac{1}{2}\lnbar^2(x) \right]
+ y \,\lnbar(y) [2 - \lnbar(x)] + 2 x \,\lnbar(x) 
\nonumber \\ &&
- \frac{5}{2} (x+y)
.
\eeq
For use below in the matching relations for gauge couplings, 
it is convenient to define a $Q$-independent combination function:
\beq
F(x,y) &\equiv& 
I(x,x,y) 
+ (x-y/2) \,\lnbar^2(x)
+ y \,\lnbar(x)\, \lnbar(y)
+ (4x-2y)\, \lnbar(x)
\nonumber \\ &&
-8x\, \lnbar(y) + [(4x-y)^2/6x] \ln(y/x)
-x/3 + 31y/6 - y^2/3x ,
\label{eq:defFxy}
\eeq
which has the nice property that the following limit is finite:
\beq
\lim_{y \rightarrow 4x} \> 
{F(x,y)}/{(y-4x)^3} &=& [2\ln(2)-1]/60 x^2 .
\label{eq:Fxylimit}
\eeq
Although the definitions in terms of ordinary 
dilogarithms are
convenient for computer numerical evaluation, it is perhaps worth nothing that 
for $y\leq 4x$, one can also write
\beq
F(x,y) &=&
(4x - y) \left [
\frac{1}{2} \Phi(y/4x) + 
\left (\frac{4}{3} + \frac{y}{6x}\right ) \ln(x/y)
 - \frac{4}{3} + \frac{y}{3x} \right ]
 ,
\eeq
where 
\beq
\Phi(z) &=& 4 \sqrt{\frac{z}{1-z}} \,{\rm Cl}_2 (2 \arcsin(\sqrt{z}))
,
\eeq
with the Clausen integral function defined by
\beq
{\rm Cl}_2(x) = -\int_0^x dt \,\ln(2 \sin(t/2)).
\eeq

\section{Decoupling relations in the Standard Model\label{sec:productiondecay}}
\setcounter{equation}{0}
\setcounter{figure}{0}
\setcounter{table}{0}
\setcounter{footnote}{1}

Consider simultaneous decoupling of 
$t$, $h$, $Z$, and $W$ from the Standard Model
at a scale $Q$. The matching relations for
the low-energy effective theory renormalizable
parameters in the $\MSbar$ scheme can be written as:
\beq
\alpha &=& \frac{e^2}{4\pi} \left [
1 + \sum_{\ell = 1}^\infty
\frac{1}{(16\pi^2)^\ell} \, \theta_\alpha^{(\ell)}
\right ]
,
\label{eq:thetaalpha}
\\
\alpha_S &=& \frac{g_3^2}{4\pi} \left [
1 +  \sum_{\ell = 1}^\infty
\frac{1}{(16\pi^2)^\ell} \, \theta_{\alpha_S}^{(\ell)}
\right ]
,
\\
m_f &=& \frac{y_f v}{\sqrt{2}} \left [1 +  \sum_{\ell = 1}^\infty
\frac{1}{(16\pi^2)^\ell}\, \theta_{m_f}^{(\ell)}
\right ],
\qquad\quad (f = b,c,s,u,d,\tau,\mu,e)
,
\label{eq:thetamf}
\eeq
where the $\ell$-loop contributions 
$\theta_X^{(\ell)}$ on the right sides are
functions of the parameters $g_3$, $g$, $g'$, $y_t$, $v$, and
the matching renormalization scale $Q$.
The effects of $y_f$ for $f\not= t$ are negligible and therefore neglected, 
except of course for the leading
factor of $y_f$ in eq.~(\ref{eq:thetamf}). Results below are
expressed\footnote{It is also easy to re-express these 
results in terms of pole squared 
masses $M_h^2$, $M_W^2$, $M_Z^2$, $M_t^2$, by  using the 
1-loop expressions relating them to $h$, $W$, $Z$, $t$
(found e.g.~in refs.~\cite{Martin:2014cxa,Martin:2015lxa,Martin:2015rea,Martin:2016xsp}, 
in the notations and VEV convention of the present paper), 
but that will not be done explicitly here.} in terms
of the running $\MSbar$ 
squared masses defined in eq.~(\ref{eq:defZ})-(\ref{eq:deft}).

\subsection{Matching of $\alpha$\label{subsec:alphamatch}}
\setcounter{footnote}{1}

\begin{figure}[!t]

\epsfxsize=0.39\linewidth
\epsffile{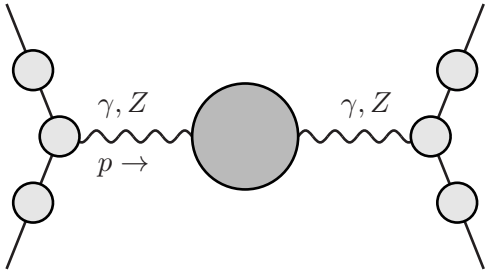}

\begin{minipage}[]{0.95\linewidth}
\caption{\label{fig:alphapole}The decoupling matching relation for $\alpha$
in the Standard Model is obtained from the residue of the 
pole at $p^2 =0$ in the amplitude for neutral current scattering of charged
particles, represented by the straight lines. By choosing the 
scattering charged particles to
be vector-like singlets under $SU(2)_L$ and to have infinitesimal $U(1)_Y$ charges,
the one-particle irreducible vertex corrections and external state
propagator corrections (depicted as the smaller light gray blobs) 
are parametrically suppressed
by an arbitrary amount, so that only
the mixed $\gamma,Z$ propagator corrections (larger, darker gray blob)
contribute.}
\end{minipage}
\end{figure}

To determine the matching condition for $\alpha$ in the 
low-energy theory, consider the residue of the pole at $p^2=0$
in the neutral current channel amplitude for scattering of charged particles,
as depicted in Figure \ref{fig:alphapole}. This is done
first in the full Standard Model including both $\gamma$ and $Z$ contributions
to the neutral current,
and then in the low-energy effective theory where only $\gamma$ exchange
contributes. Requiring that the results of the two calculations agree gives the matching condition. 

In order to avoid complications involving
charged particle propagator and vertex corrections, 
it is convenient to use a trick, 
by taking the charged particles to be vector-like
singlets under weak 
isospin $SU(2)_L$ and
to carry infinitesimal electric charges, which are therefore also equal to their
$U(1)_Y$ charges. This ensures that the one-particle-irreducible
vertex corrections and the charged particle propagator corrections
to the amplitude are parametrically suppressed 
by an arbitrary amount relative to the bosonic propagator corrections, 
due to higher powers of the infinitesimal charges, 
and so can be neglected. The result for the
matching of the electromagnetic coupling then follows 
only from consideration of the corrections to the $\gamma,Z$ system propagator. 
The idea behind this trick is 
that $U(1)_{\rm EM}$ gauge invariance guarantees that the matching 
condition for the electromagnetic coupling cannot 
depend on the quantum numbers of the
charged states to which the neutral current couples, so the same result 
must obtain for other scattering processes involving chiral fermions including
$SU(2)_L$ doublets such as those in the Standard Model, where vertex and fermion propagator corrections are non-trivial.

The propagator matrix for the $\gamma, Z$ system 
can be written in terms of the components of the (transverse) 
one-particle-irreducible self-energy functions  
$\Pi_{ab}(s)$ for $a,b = \gamma,Z$ and $s = -p^2$, as 
$iG (\eta^{\mu\nu} - p^\mu p^\nu/p^2)$, where
\beq
G^{-1} &=& \begin{pmatrix}
s - \Pi_{\gamma\gamma}(s) & -\Pi_{\gamma Z}(s) \\
-\Pi_{\gamma Z}(s) & \phantom{xx}s - m_Z^2 - \Pi_{ZZ}(s)
\end{pmatrix}
.
\eeq
It follows that
\beq
G_{\gamma\gamma} &=& \frac{1}{s - \widetilde\Pi_{\gamma\gamma}} 
\>
,
\\
G_{\gamma Z} &=& \frac{\Pi_{\gamma Z}}{
\bigl (s - m_Z^2 - \Pi_{ZZ} \bigr )
\bigl (s - \widetilde\Pi_{\gamma\gamma} \bigr ) }
\>
,
\\
G_{ZZ} &=& 
\frac{s - \Pi_{\gamma\gamma}}{
\bigl (s - m_Z^2 - \Pi_{ZZ})
\bigl (s - \widetilde\Pi_{\gamma\gamma} \bigr ) 
}
\>
,
\eeq
where
\beq
\widetilde \Pi_{\gamma\gamma} &\equiv & \Pi_{\gamma\gamma} + 
\left (\Pi_{\gamma Z}\right )^2/(s - m_Z^2 - \Pi_{ZZ}).
\eeq

Now the neutral current interaction amplitude between two 
$SU(2)_L$ singlet states with infinitesimal $U(1)_Y$ charges
is just proportional to
\beq
{\cal A} &=& g^{\prime 2} G_{YY}
\>=\>
g^{\prime 2}
\left [c_W^2 G_{\gamma\gamma} - 2 c_W s_W G_{\gamma Z} + s_W^2 G_{ZZ}
\right ]
,
\eeq
where $c_W = g/\sqrt{g^2 + g^{\prime 2}}$ and 
$s_W = g'/\sqrt{g^2 + g^{\prime 2}}$. It follows that
\beq
{\cal A}
&=& 
e^2 
\left [1  + \frac{(g'/g)^2 (s - \Pi_{\gamma\gamma}) - (2 g'/g)  \Pi_{\gamma Z} 
}{s - m_Z^2 - \Pi_{ZZ}} \right]
/\left [s - \widetilde \Pi_{\gamma\gamma} \right ]
.
\eeq
The existence of a massless photon pole in the amplitude at $s=0$ therefore implies 
\beq
\widetilde \Pi_{\gamma\gamma}(0) = 0,
\label{eq:spole}
\eeq
and the residue of the pole in ${\cal A}$ 
is a gauge-invariant physical observable:
\beq
{\rm res}({\cal A}) &=&
e^2 \left [1 + \frac{(g'/g) \,\Pi_{\gamma Z}(0)}{m_Z^2 + \Pi_{ZZ}(0)}\right]^2 /
\left [1 - \widetilde \Pi_{\gamma\gamma}'(0)\right ] .
\eeq
Here eq.~(\ref{eq:spole}) has been used to eliminate $ \Pi_{\gamma\gamma}(0)$ from the numerator.
So far no perturbative expansions or approximations or assumptions about the particular choice of gauge-fixing scheme have been used.

The calculation of the $\Pi_{ab}(s)$ is then performed in Landau gauge 
in a loop expansion in terms of bare parameters (with no counterterm diagrams)
in $d = 4 - 2 \epsilon$ dimensions, and at the end the result 
for the residue of the pole, ${\rm res}({\cal A})$, 
is translated to the $\MSbar$ scheme by 
the standard parameter redefinitions
that give bare parameters
(including the VEV) in terms of $\MSbar$ ones.\footnote{In 
the same notations and conventions as the present paper, they can be found in
eqs.~(2.5)-(2.24) of ref.~\cite{Martin:2014cxa},
eqs.~(2.3)-(2.12) of ref.~\cite{Martin:2015lxa},
and eqs.~(2.5)-(2.8) of ref.~\cite{Martin:2015rea}.}  
This procedure is simpler and easier than 
using separate counterterm Feynman rules from the start, and the cancellation
of poles in $\epsilon$ provides a check. The verification of
eq.~(\ref{eq:spole}) through 2-loop order gives another check.

The calculation of ${\rm res}({\cal A})$ is then repeated in the low-energy 
theory with $t, h, Z, W$ absent, 
and therefore no $\Pi_{\gamma Z}$ or $\Pi_{ZZ}$, so that 
$\Pi_{\gamma\gamma}(0) = 0$, and ${\rm res}({\cal A}) = 
4 \pi \alpha/[1 - \Pi_{\gamma\gamma}'(0)]$.
Requiring that the two results 
for the observable ${\rm res}({\cal A})$
are equal gives the matching condition for
the electromagnetic coupling, after taking into account the 1-loop
matching for the light fermion 
masses between the two theories, from subsection
\ref{subsec:matchmf} below. Note that non-perturbative effects from confined
light quarks
are common to the two versions of ${\rm res}({\cal A})$, and so cancel out.

At 1-loop order, one obtains the well-known result:
\beq
\theta_\alpha^{(1)} &=&  e^2 \left [
\frac{2}{3} - 7\, \lnbar(W) + \frac{16}{9}\, \lnbar(t) \right] .
\label{eq:thetaalpha1}
\eeq
For the 2-loop contribution to the matching, I obtain:
\beq
\theta_\alpha^{(2)} &=&  
e^2 g_3^2 \left [
-\frac{64}{9}\, \lnbar(t) - \frac{208}{27} \right ]
+ e^2 y_t^2 
\biggl \{
\frac{16 t (h-t)}{3 h(4t-h)^2} F(t,h) - \frac{16 t}{9 h} [1 + \ln(h/t)]
\nonumber \\ &&
+ \frac{16 t}{9 (4t-Z)^3}
 \left [ t \left (80 W/Z - 7 - 64 W^2/Z^2 \right ) + 8 Z - 40 W + 32 W^2/Z \right ] F(t,Z)
\nonumber \\ &&
+ 4 \Bigl [
\left (80 W/Z - 64 W^2/Z^2 \right ) \left [1 + \ln(Z/t)\right ]
+ 2 + 3\, \lnbar(h) - 7\, \lnbar(Z) - 14\, \lnbar(t)
\Bigr ]/27
\nonumber \\ &&
+ 22\, \lnbar(t) -\frac{43}{4}
\biggr \}
+ e^2 g^2 \biggl \{ 
\frac{3 W (3 h^2 - 12 h W + 4 W^2)}{h (4W-h)^3} F(W,h)
+ \left (
\frac{W}{h} - 2 
\right )\, \lnbar(h)
\nonumber \\ &&
+ \frac{9 W (4 W^2 - 4 W Z + 3 Z^2)}{Z^2 (4W-Z)^2} 
F(W,Z)
+ \left (
\frac{661 Z}{108 W}
- \frac{491}{27}
+ \frac{319 W}{27 Z} 
+ \frac{12 W^2}{Z^2}
\right ) \, \lnbar(Z)
\nonumber \\ &&
+ \left (
\frac{20}{3} + \frac{37 W}{3Z} - \frac{12 W^2}{Z^2} - \frac{W}{h}
\right ) \, \lnbar(W)
+ \frac{5 t}{3 (t-W)} \ln(t/W)
\nonumber \\ &&
+ \frac{31}{81}
- \frac{3 h}{4 W}
+ \frac{W}{h}
+ \frac{12 W^2}{Z^2}
- \frac{799 W}{27 Z}
- \frac{1057 Z}{324 W}
\biggr \}
\nonumber \\ &&
+ e^4 \Bigl \{
49 \, \lnbar^2(W)
- \frac{224}{9} \, \lnbar(t) \, \lnbar(W)
+ \frac{256}{81} \, \lnbar^2(t)
\Bigr \}.
\label{eq:thetaalpha2}
\eeq
The $g_3^2$ part of eq.~(\ref{eq:thetaalpha2}) can be checked to be consistent 
with previously known results for the relation between the Thomson scattering
value of $\alpha$ and its $\MSbar$ version, e.g.~in ref.~\cite{Kniehl:2015nwa}.

The presentation of eq.~(\ref{eq:thetaalpha2}) is 
made simpler by the use of the function $F(x,y)$ defined in 
eq.~(\ref{eq:defFxy}) above.
Equation (\ref{eq:Fxylimit}) shows immediately 
that $\theta^{(2)}_\alpha$
is finite and well-defined for $h = 4t$ and $Z = 4 t$ and
$h = 4 W$ and $Z = 4W$ as well as $t=W$, 
despite the presence of denominators proportional to 
$(4t-h)^2$ and $(4t-Z)^3$ and $(4W-h)^3$ and $(4W-Z)^2$ and $t-W$
in eq.~(\ref{eq:thetaalpha2}). 
This is a useful check, since there is no physical
reason why anything untoward should happen at these special cases, 
even though of course none of them are  
close to being realized in our world. Additional checks are provided
by the absence of poles $1/\epsilon$ and $1/\epsilon^2$,
and by the cancellation\footnote{More generally, 
in the Landau-gauge tadpole-free scheme this check 
is a non-trivial counterpart to the gauge-invariance
check that one would obtain by instead working in a general 
gauge fixing  with a tree-level VEV and non-vanishing tadpoles.}
of dependence on the Landau gauge Goldstone boson squared 
mass, after Goldstone boson resummation \cite{Martin:2014bca,Elias-Miro:2014pca}.
I have further checked that renormalization
group invariance is satisfied by eqs.~(\ref{eq:thetaalpha}),
(\ref{eq:thetaalpha1}), and (\ref{eq:thetaalpha2}), by computing
the $Q$ derivative of each side using the known beta functions of
the low-energy and high-energy theories and the direct
$Q$ dependence of the function $\lnbar(x)$. 
[Note that $F(x,y)$ has no $Q$ dependence.] 
In principle, this check should be merely equivalent to 
requiring the absence of poles in $\epsilon$, but in practice 
it also checks intermediate steps in the calculations.

\subsection{Matching of $\alpha_S$\label{subsec:alphaSmatch}}
\setcounter{footnote}{1}

For the decoupling relation of $\alpha_S$, the result has already been
obtained in pure QCD to 
1-loop \cite{Weinberg:1980wa}, 
2-loop \cite{Bernreuther:1981sg,Larin:1994va},
3-loop \cite{Chetyrkin:1997un},
and 4-loop order \cite{Schroder:2005hy,Chetyrkin:2005ia}, and
at complete 2-loop order by Bednyakov \cite{Bednyakov:2014fua}. 
I have re-calculated the latter result,
finding complete agreement:
\beq
\theta_{\alpha_S}^{(1)} &=&  \frac{2}{3} g_3^2 \, \lnbar(t) 
,
\label{eq:thetaalphaS1}
\\
\theta_{\alpha_S}^{(2)} &=&  
g_3^4 \left [ \frac{22}{9} + \frac{22}{3}\, \lnbar(t) + \frac{4}{9}\, \lnbar^2(t)
\right ]
\,+\, g_3^2 y_t^2 \biggl \{
\frac{2 t (h-t)}{h(4t-h)^2} F(t,h) - \frac{2 t}{3 h} \left [1 + \ln(h/t) \right ]
\nonumber \\ &&
+ \frac{2 t}{3 (4t-Z)^3} \left [ t \left (80 W/Z - 7 - 64 W^2/Z^2 \right ) + 8 Z - 40 W + 32 W^2/Z \right ] F(t,Z)
\nonumber \\ &&
+ \Bigl [
\left (80 W/Z - 64 W^2/Z^2 \right ) \left [1 + \ln(Z/t)\right ]
+ 2 + 3\, \lnbar(h) - 7\, \lnbar(Z) - 14\, \lnbar(t)
\Bigr ]/18 \biggr \} 
\nonumber \\ &&
+ g_3^2 g^2 \biggl \{
\frac{8(W-Z)}{9Z} \lnbar(t)
+ 3\, \lnbar(W)
+ \left (\frac{25 Z}{18 W} -\frac{13}{9} + \frac{14 W}{9 Z} \right ) \lnbar(Z)
\nonumber \\ &&
+ \frac{t}{t-W} \ln(t/W)
- \frac{49}{27} - \frac{W}{18 Z} - \frac{163 Z}{216 W}
\biggr \} .
\label{eq:thetaalphaS2}
\eeq
Although equivalent, 
the presentation in eq.~(\ref{eq:thetaalphaS2}) is somewhat more compact
than the expression given in ref.~\cite{Bednyakov:2014fua}.
This is due in part to the use of the function $F(x,y)$ defined 
in eq.~(\ref{eq:defFxy}) above, and also because the results are 
given in terms of running $\MSbar$ 
squared masses here, rather than pole masses as in 
ref.~\cite{Bednyakov:2014fua}; converting to the top-quark pole mass 
in eq.~(\ref{eq:thetaalphaS1}) just contributes some additional 2-loop terms
involving the 1-loop top-quark on-shell self-energy.
  
The pure QCD contributions to decoupling the top quark
at 
3-loop \cite{Chetyrkin:1997un}
and 4-loop 
\cite{Schroder:2005hy,Chetyrkin:2005ia} order 
are also reproduced here for the sake of completeness:
\beq
\theta_{\alpha_S}^{(3)} &=&  g_3^6  \left [
\frac{8}{27} \,\lnbar^3(t) 
-3 \,\lnbar^2(t)  
+ \frac{620}{9}\, \lnbar(t) 
+ 35.123151
\right ]
,
\label{eq:thetaalphaS3}
\\
\theta_{\alpha_S}^{(4)} &=&  g_3^8  
\left [
\frac{16}{81} \, \lnbar^4(t)
+ \frac{4706}{81} \, \lnbar^3(t) 
- \frac{1231}{27}\, \lnbar^2(t)  
+ 245.856958\, \lnbar(t) 
- 109.765121 
\right ]
.
\phantom{xx}
\label{eq:thetaalphaS4}
\eeq
The coefficients involving 
irrational numbers (available in their full glory in 
refs.~\cite{Chetyrkin:1997un,Schroder:2005hy,Chetyrkin:2005ia,Kniehl:2006bg}) 
have been reduced to decimal approximations here and in similar expressions below, 
for the sake of brevity.

\subsection{Matching of running fermion masses\label{subsec:matchmf}}
\setcounter{footnote}{1}

Now consider the decoupling relations for the masses of the fermions lighter
than the top quark. The matching functions can be given generically for
fermions other than the bottom quark, which is different because it has
a direct coupling to the top quark and $W$ boson. For a generic fermion,
\beq
(Q_f, I_3^f, C_f) = 
\left \{ \begin{array}{rrrl}
(2/3,\> &1/2,\>\, &4/3) & \qquad(f = t,c,u),
\\[-5pt]
(-1/3,\> &-1/2,\>\, &4/3) & \qquad(f = b,s,d),
\\[-5pt]
(-1,\> &-1/2,\>\, &0) & \qquad(f = \tau,\mu,e),
\end{array}
\right.
\label{eq:QIC}
\eeq
are the notations for electric charge $Q_f$, $I_3^f$ for the third component
of weak isospin of the left-handed fermion, and 
$C_f$ for the $SU(3)_c$ Casimir invariant.

The method used is to require equality between two computations 
of the pole mass for each  light fermion, 
first in the full Standard Model theory and then again in the low-energy 
effective theory without $t,h,Z,W$. 
The strategy and details of the calculation of the light 
fermion pole masses that I have used 
are very similar to those described already in ref.~\cite{Martin:2016xsp}
for the top quark, and so will not be reviewed here.

The resulting 1-loop order threshold corrections to the light fermion
masses are:
\beq
\theta_{m_f}^{(1)} &=& 
\frac{9 g^2 + 3 g^{\prime 2}}{16}
+ Q_f g^{\prime 2} \left [I_3^f + Q_f (W/Z - 1) \right ]
\left [3\, \lnbar(Z) - 5/2 \right ] 
,
\label{eq:theta1mf}
\\
\theta_{m_b}^{(1)} &=& \theta_{m_d}^{(1)} + \frac{3}{4} y_t^2 \left [
\frac{5}{6} - \lnbar(t) 
+
\left (\frac{W}{t-W}\right )^2 \ln(t/W) - \frac{W}{t-W} \right ],
\label{eq:theta1mb}
\eeq
for a generic fermion other than the bottom quark, and for the bottom quark,
respectively. In the case of the bottom quark, only the leading order
in an expansion in $y_b^2$ has been kept. The next term in the
expansion is
\beq
\Delta\theta_{m_b}^{(1)} &=& y_b^2
\biggl [
\frac{3}{4} \, \lnbar(h)
+ \frac{1}{4}\, \lnbar(t)
+ \frac{W^2 (3 t^2 + 4 t W - W^2)}{4 (t-W)^4} \,\ln(t/W)
- \frac{W^2 (7t-W)}{4 (t-W)^3}
\nonumber \\ &&
+ \frac{7 Z^2 + 16 W Z - 32 W^2}{36 Z^2} \, \lnbar(Z)
- \frac{4 (Z-W)(2W+Z)}{9 Z^2} \, \lnbar(b)
\nonumber \\ &&
- \frac{91}{216} - \frac{8 W}{27 Z} + \frac{16 W^2}{27 Z^2}
\biggr ].
\eeq
However, since $y_b^2/16\pi^2 \><\> 2 \times 10^{-6}$, this contribution
is negligible.

The 2-loop order threshold function for the bottom quark mass takes the form:
\beq
\theta_{m_b}^{(2)} &=& 
\frac{4}{3} g_3^4 \left [ \lnbar^2(t) + \frac{5}{3} \lnbar(t) + \frac{89}{36} \right ]
+ g_3^2 y_t^2 \Bigl [
(8 t^2 - 8 t W + 6 W^2) I(0,t,W)
\nonumber \\ &&
+ t (7 t^2 - 17 t W + 22 W^2)\, \lnbar^2(t)
+ 2 t W (4t-7W) \, \lnbar(t) \, \lnbar(W)
\nonumber \\ &&
+ (35 t^2 W - 23 t^3 - 56 t W^2 + 16 W^3) \, \lnbar(t)
- (2t-3W) (7t-3W) W \lnbar(W)
\nonumber \\ &&
+ 92 t^3/3 - 19 t^2 W + 17 t W^2 + 4 W^3/3
\Bigr ]/(t-W)^3
\nonumber \\ &&
+ \frac{4}{3} g_3^2 \biggl \{
\frac{g^{\prime 2}}{6}  (1 + 2 W/Z) \left [\lnbar(Z) - 17/12 \right ]
+ \frac{9}{4} g^2 \lnbar(W)
\nonumber \\ &&
+ \frac{9}{8} (g^2 + g^{\prime 2}) \lnbar(Z)
- \frac{15}{32} (3 g^2 + g^{\prime 2})
\biggr \}
\nonumber \\ &&
+ \sum_{j=1}^{12} b^{(2)}_j {\cal I}^{(2)}_j
+ \sum_{j=1}^4\sum_{k=1}^j b^{(1,1)}_{j,k} {\cal I}^{(1)}_j {\cal I}^{(1)}_k
+ \sum_{j=1}^4 b^{(1)}_{j} {\cal I}^{(1)}_j 
+ b^{(0)}
.
\label{eq:thetam2b}
\eeq
The part that does not contain the strong coupling $g_3$ involves
coefficients of 2-loop integral functions and logarithms
from the lists
\beq
{\cal I}^{(2)} &=& \bigl \{
\zeta_2,\>\> I(0,h,W),\>\> I(0,h,Z),\>\> I(0,W,Z),\>\> 
I(h,W,W),\>\> I(h,Z,Z),\>\> 
\nonumber \\ && 
I(t,t,Z),\>\> I(W,W,Z),\>\> I(0,t,W),\>\> I(h,t,t),\>\> I(h,t,W),\>\> I(t,W,Z)
\bigr \} 
\label{eq:defI2basis}
,
\\
{\cal I}^{(1)} &=& \{\lnbar(t),\>\> \lnbar(h),\>\> \lnbar(Z),\>\> \lnbar(W) \},
\eeq
respectively. It cannot be simplified to a length reasonable for printing, 
and so is not given explicitly above in its full form, but
instead in an ancillary electronic file distributed with this paper,
called {\tt theta2mb}. The individual coefficients 
$b^{(2)}_j$, $b^{(1,1)}_{j,k}$, $b^{(1)}_{j}$, and $b^{(0)}$ are rational
functions of the input parameters $t$, $h$, $Z$, $W$, and $v$. 
Many of them have 
poles in one or more of 
the quantities $t-W$ and $4W-h$ and $4Z-h$ and $4t-Z$ and
$\lambda(t,W,Z)$ and $\lambda(t,W,h)$, but I have checked that 
the total
$\theta_{m_b}^{(2)}$ is nevertheless finite when each of these 
quantities vanishes. The format used in the 
ancillary file {\tt theta2mb} is compatible with
inclusion in computer code for easy numerical evaluation using
eqs.~(\ref{eq:deflnbar})-(\ref{eq:defFxy}).
Additional checks follow, as usual, from the
absence of poles $1/\epsilon^2$ and $1/\epsilon$ upon
translating to the $\MSbar$ scheme, and by the cancellation of
contributions involving the Landau gauge Goldstone boson mass.

For generic fermions $f = (c,s,u,d,\tau,\mu,e)$, the 2-loop threshold
functions are similarly found to be:
\beq
\theta_{m_f}^{(2)} &=& 
C_f g_3^4 \left [ \lnbar^2(t) + \frac{5}{3} \lnbar(t) + \frac{89}{36} \right ]
+ C_f g_3^2 \biggl \{
3 g^{\prime 2} Q_f 
\left [I_3^f + Q_f (W/Z - 1) \right ]
\left [\lnbar(Z) - 17/12 \right ]
\nonumber \\ &&
+ \frac{9}{4} g^2 \lnbar(W)
+ \frac{9}{8} (g^2 + g^{\prime 2}) \lnbar(Z)
- \frac{15}{32} (3 g^2 + g^{\prime 2})
\biggr \}
\nonumber \\ &&
+ \sum_{j=1}^8 c^{(2)}_j {\cal I}^{(2)}_j
+ \sum_{j=1}^4\sum_{k=1}^j c^{(1,1)}_{j,k} {\cal I}^{(1)}_j {\cal I}^{(1)}_k
+ \sum_{j=1}^4 c^{(1)}_{j} {\cal I}^{(1)}_j 
+ c^{(0)}
,
\label{eq:theta2mf}
\eeq
where the contributions independent of $g_3$ involve coefficients that
are again too complicated to show in print, 
and so are relegated to an electronic file called {\tt theta2mf}
distributed as an ancillary to this paper. 
Note that the last four functions in the list eq.~(\ref{eq:defI2basis}) 
do not appear in eq.~(\ref{eq:theta2mf}). 
The individual coefficients 
$c^{(2)}_j$, $c^{(1,1)}_{j,k}$, $c^{(1)}_{j}$, and $c^{(0)}$
are again rational functions of $t$, $h$, $Z$, $W$, and $v$, with pole
singularities at $4Z-h$ and $4t-Z$, but the total is free of these singularities.

The pure QCD threshold corrections for light quark masses 
were already known up to 3-loop order from Chetyrkin, Kniehl, and Steinhauser
in ref.~\cite{Chetyrkin:1997un}
and Liu and Steinhauser at 4-loop order in ref.~\cite{Liu:2015fxa}.
They are listed here for the sake of completeness. For each 
quark $q = (b,c,s,u,d)$:
\beq
\theta_{m_q}^{(3)} &=& g_3^6 \left [ 
-\frac{152}{27} \, \lnbar^3(t) 
+ \frac{700}{27} \, \lnbar^2(t)
+ 111.047973 \, \lnbar(t)
+ 126.160947
\right ]
,
\label{eq:thetamq3}
\\
\theta_{m_q}^{(4)} &=& g_3^8 \biggl [ 
\frac{830}{27} \, \lnbar^4(t) 
-\frac{10984}{81} \, \lnbar^3(t) 
- 543.379386 \, \lnbar^2(t) 
+ 452.388432 \, \lnbar(t) 
\nonumber \\ &&
+ 236.908052
\biggr ]
.
\label{eq:thetamq4}
\eeq
Note that the preceding equations apply specifically to the 
decoupling of the top quark from the theory. Again the known 
irrational parts have been replaced by decimal approximations.


\section{Decoupling of lighter fermions in the QCD+QED effective theory\label{sec:matchingeff}}
\setcounter{equation}{0}
\setcounter{figure}{0}
\setcounter{table}{0}
\setcounter{footnote}{1}

In this section, I provide the decoupling relations appropriate for further
sequential decoupling of fermions within the QCD+QED theory. 
None of the results in this section are new, as the QCD parts of
these are now well-known, and the QED contributions at up to 2-loop order 
and certain light mass expansions can 
be easily inferred from those found in the existing literature. 
They are collected here for the sake of completeness.

The notation adopted here assumes that 
a generic fermion, denoted $F$, is to be 
decoupled.\footnote{In the Standard Model, the formulas below are not practically
applicable with $F = u,d,s$, 
because QCD perturbation theory is not under control. Instead, the RPP 
\cite{Tanabashi:2018oca}
quotes the $\MSbar$ masses at $Q=2$ GeV.}
The charge and QCD Casimir 
quantum numbers of $F$ are to be denoted $Q_F$ and $C_F$ respectively,
just as in eq.~(\ref{eq:QIC}), and the index $T_F$ equals $1/2$ when the
decoupled fermion $F$ is a quark, and is 0 if it is a lepton, while the
number of colors $N_F$ is 3 when $F$ is a quark and 1 when $F$ is a lepton.
The decoupling scale $Q$ associated with the matching of parameters
is again arbitrary, but typically should be chosen to be comparable
to the mass of $F$, in order to avoid large logarithms
in observables calculated after using the renormalization group
equations to run the surviving parameters to lower energies. The 
running $\MSbar$ parameters of the high-energy 
$SU(3)_c \times U(1)_{\rm EM}$ theory will be denoted $\alpha$, $\alpha_S$,
$F \equiv m_F^2$, and $m_f$, 
where $f$ runs over the list of the lighter fermions which
are not being decoupled. For the low-energy 
theory with $F$ decoupled,
the parameters are distinguished by an underline, so they are 
$\underline{\alpha}$, $\underline\alpha_S$, and $\underline m_f$.
The number of light quark flavors among the fermions $f$ in the decoupled theory
(which will also include leptons) will be denoted $n_q$.

The decoupling relations can then be written in the form:
\beq
\underline\alpha(Q) &=& \alpha(Q) \left [ 
1 + \sum_{\ell = 1}^\infty \frac{1}{(4\pi)^\ell} \vartheta_\alpha^{(\ell)} \right]
,
\\
\underline\alpha_S(Q) &=& \alpha_S(Q) \left [ 
1 + \sum_{\ell = 1}^\infty \frac{1}{(4\pi)^\ell} \vartheta_{\alpha_S}^{(\ell)} \right]
,
\\
\underline{m}_f(Q) &=& m_f(Q) \left [ 
1 + \sum_{\ell = 1}^\infty \frac{1}{(4\pi)^\ell} \vartheta_{m_f}^{(\ell)} \right]
\qquad\quad (f\not= F).
\eeq
(Note that the symbol $\vartheta$ is used to denote the threshold corrections
within the QCD+QED theory in this section,
in distinction with the symbol $\theta$ used in the previous section for decoupling $t,h,Z,W$.)
Then for the matching coefficients for the electromagnetic coupling, one has
at the scale $Q$ where $F$ is decoupled:
\beq
\vartheta_\alpha^{(1)} &=& \frac{4}{3} N_F Q_F^2 \alpha \> \lnbar(F)
,
\\
\vartheta_\alpha^{(2)} &=& \left [\frac{4}{3} N_F Q_F^2 \alpha \>\lnbar(F)\right ]^2 
- N_F Q_F^2 \alpha 
\left (  C_F \alpha_S + Q_F^2 \alpha \right )
\left [4\, \lnbar(F) + \frac{13}{3} \right ]
.
\eeq
For the QCD coupling, the results through 2-loop order including QED effects
are:
\beq
\vartheta_{\alpha_S}^{(1)} &=& \frac{4}{3} T_F \alpha_S \> \lnbar(F)
,
\\
\vartheta_{\alpha_S}^{(2)} &=& 
\left [ \frac{4}{3} T_F \alpha_S \> \lnbar(F) \right ]^2
- T_F \alpha_S \left (C_F \alpha_S + Q_F^2 \alpha\right )
\left [4\,\lnbar(F) + \frac{13}{3} \right ] 
\nonumber \\ &&
+ T_F C_A \alpha_S^2 \left [
\frac{20}{3} \lnbar(F) + \frac{32}{9} \right ]
,
\eeq
where $T_F = 1/2$ when $F$ is a quark, and $T_F= 0$ 
when $F$ is a lepton, and $C_A = 3$.
The pure QCD contributions at 3-loop and 4-loop order, which  
apply only if $F$ is a quark, are found from
refs.~\cite{Chetyrkin:1997un} and \cite{Schroder:2005hy,Chetyrkin:2005ia}:
\beq
\vartheta_{\alpha_S}^{(3)} &=& \alpha_S^3
\biggl [
\frac{8}{27}\, \lnbar^3(F) + 
\left ( \frac{53}{9} - \frac{16}{9} n_q \right ) \lnbar^2(F)
+ \left (\frac{955}{9} - \frac{67}{9} n_q \right ) \lnbar(F)
\nonumber \\ && 
+ 62.211628
- \frac{2633}{486} n_q
\biggr ]
,
\\
\vartheta_{\alpha_S}^{(4)} &=& \alpha_S^4
\biggl [
\frac{16}{81}\, \lnbar^4 (F) \,+\,
\left (
\frac{3766}{81} + \frac{508}{81} n_q - \frac{64}{81} n_q^2
\right ) \, \lnbar^3 (F) \nonumber \\ &&
+
\left (
\frac{4354}{27} - \frac{2966 }{81} n_q - 
\frac{77 }{81} n_q^2 \right ) \, \lnbar^2(F) 
\nonumber \\ &&
+
\left (
2157.863053 - 335.316171 n_q 
- \frac{6865}{729} n_q^2 \right ) \, \lnbar(F) 
\nonumber \\ &&
+ 1323.608830 
- 258.542470 n_q 
- 5.626464 n_q^2
\biggr ]
.
\eeq
These can be used with $n_q = 4$ when $F$ is the bottom quark, and $n_q = 3$
when $F$ is the charm quark. The formulas with $n_q=5$ of course coincide 
with that for decoupling 
the top quark, as in eqs.~(\ref{eq:thetamq3})-(\ref{eq:thetamq4}) above.

The 1-loop and 2-loop
threshold corrections for each light fermion mass $m_f$ when decoupling the
fermion $F$ in the $SU(3)_c \times U(1)_{\rm EM}$ theory are:
\beq
\vartheta_{m_f}^{(1)} &=& 0,
\\
\vartheta_{m_f}^{(2)} &=& 2 \left (T_F C_f \alpha_S^2
+ N_F Q_F^2 Q_f^2 \alpha^2 \right )
\left [\lnbar^2(F) + \frac{5}{3} \lnbar(F) + \frac{89}{36}  + \Delta_2(f/F)\right ]
,
\eeq
where the last term is the power-suppressed mass correction, with
$f,F$ being the $\MSbar$ squared masses and 
\beq
\Delta_2(r) &=&
r \left (\frac{8}{15} \ln(r) - \frac{76}{75}\right )
+ r^2 \left ( \frac{9}{70} \ln(r) - \frac{1389}{9800} \right)
+ {\cal O}(r^3)
.
\eeq
This effect
is mentioned because the squared mass ratios occurring in the decoupling 
of the light fermions (notably, $c/b \sim 0.1$) are not quite as suppressed as 
$b/t$ in the decoupling of the top quark in the previous section, but its numerical impact is still quite small.
It can be obtained from the 2-loop result for a quark pole mass in the presence of
other massive and massless quarks, in ref.~\cite{Gray:1990yh}.
The pure QCD corrections are also known at 3-loop and 4-loop orders
from refs.~\cite{Chetyrkin:1997un} and \cite{Liu:2015fxa} respectively:
\beq
\vartheta_{m_f}^{(3)} &=&
\alpha_S^3 \biggl [
\left (\frac{16}{27} n_q - \frac{232}{27} \right ) \lnbar^3(F)
+ \frac{700}{27}\, \lnbar^2(F)
+ \left ( \frac{212}{27} n_q +  71.788714 \right ) \lnbar(F)
\nonumber \\ && 
+ 118.248112 + 1.582567 n_q
+ \Delta_3(f/F)
\biggr ]
,
\\
\vartheta_{m_f}^{(4)} &=&
\alpha_S^4 \biggl [
\left (\frac{8}{27} n_q^2 - \frac{80}{9} n_q + \frac{610}{9} \right )\, \lnbar^4(F)
+ \left (\frac{184}{9} n_q - \frac{19264}{81} \right )\, \lnbar^3(F)
\nonumber \\ && 
+ \left ( \frac{496}{81} n_q^2 - \frac{15650}{81} n_q  +  269.583577 \right ) \,\lnbar^2(F)
\nonumber \\ && 
+ \bigl (286.364218 + 39.625147 n_q - 1.284061 n_q^2 \bigr ) \, \lnbar(F)
\nonumber \\ &&
+ 14.375890 n_q^2 - 375.221169 n_q + 1753.616640 
\biggr ]
.
\eeq
In the 3-loop part, the small mass correction is
\beq
\Delta_3(r) &=& \frac{8}{9} (2 n_q - 31)\, \lnbar(F) \Delta_2(r) 
\nonumber \\ &&
\!\!\!\!\!\!\!\!\!\!\!\!\!
+
r \biggl  \{ \left (\frac{64}{135} n_q - \frac{451}{81} \right ) \ln^2(r) 
+ \left (\frac{84887}{7290} - \frac{128}{135} n_q \right ) \ln(r)
+ 2.77670 - 0.22452 n_q
\biggr \}
\nonumber \\ &&
\!\!\!\!\!\!\!\!\!\!\!\!\!
+ r^2 \biggl  \{ 
\left (\frac{4}{35} n_q - \frac{239}{270} \right ) \ln^2(r) 
+ \left (\frac{580157}{396900} - \frac{6}{35} n_q\right ) \ln(r)
+ 0.52092 + 0.03556 n_q
\biggr \}
\nonumber \\ &&
\!\!\!\!\!\!\!\!\!\!\!\!\!
+ {\cal O}(r^3)
,
\eeq
which can be gleaned from the
expansion of the pole mass given in ref.~\cite{Bednyakov:2016onn} 
based on the results in \cite{Bekavac:2007tk,Bekavac:2009gz}.
In the 4-loop part, the expansion is not known beyond the lowest order in 
$r = f/F$.

In applications of the above formulas, the renormalization group running between
scales requires the beta functions for the two gauge couplings and the fermion
masses, which are known in the $SU(3)_c \times U(1)_{\rm EM}$ theory at full
3-loop order including electromagnetic effects; see for example 
ref.~\cite{Bednyakov:2016onn} 
(and the Appendix of ref.~\cite{Martin:2006ub}
for a general product gauge group with an arbitrary reducible fermion
representation). The higher-order QCD corrections
to the beta function for $\alpha$ are given in ref.~\cite{Erler:1998sy}
at order $\alpha^2 \alpha_S^3$ and 
in ref.~\cite{Baikov:2012zm} at order $\alpha^2 \alpha_S^4$.
The 4-loop and 5-loop pure QCD contributions to the $\alpha_S$ beta function are
found in refs.~\cite{vanRitbergen:1997va,Czakon:2004bu} and
\cite{Baikov:2016tgj,Herzog:2017ohr}, respectively.
The 3-loop, 4-loop and 5-loop pure QCD contributions 
to the quark mass betas functions
are in \cite{Tarasov}, \cite{Chetyrkin:1997dh,Vermaseren:1997fq}, 
and \cite{Baikov:2014qja}.
Also useful in this context are the fermion pole masses, which 
are given for a general product gauge group with an arbitrary reducible fermion
representation (but assuming just one non-zero fermion mass) in the Appendix of 
ref.~\cite{Martin:2006ub}, with 4-loop pure QCD contributions 
in refs.~\cite{Marquard:2015qpa,Marquard:2016dcn}. In the case of more 
than one non-zero quark mass, expansions for small and large mass ratios 
in the 3-loop pole masses have been given in 
refs.~\cite{Bekavac:2007tk} and \cite{Bednyakov:2016onn}.

\section{Numerical results\label{sec:numerical}}
\setcounter{equation}{0}
\setcounter{figure}{0}
\setcounter{table}{0}
\setcounter{footnote}{1}

In this section, I will illustrate the numerical impact of the matching 
conditions, concentrating on the new results of this paper, 
i.e.~the shifts in the electromagnetic coupling and the light fermion masses
from decoupling $t,h,Z,W$ in the Standard Model, as a function of the
matching scale $Q$. 
For a benchmark model, I consider the 
following numerical values for Standard Model parameters at 
a reference scale $Q_0 = 173.34$ GeV:
\beq
g_3 &=& 1.1666
,
\\
g &=& 0.647550
,
\\
g' &=& 0.358521
,
\\
y_t &=& 0.93690
,
\\
\lambda &=& 0.12597,
\\
v &=& 246.647\>{\rm GeV} .
\eeq
These are then run to a matching scale 80 GeV $<Q<$ 180 GeV, 
and the figures below show the resulting matching corrections obtained 
in subsections \ref{subsec:alphamatch} and \ref{subsec:matchmf}.

\begin{figure}[!t]
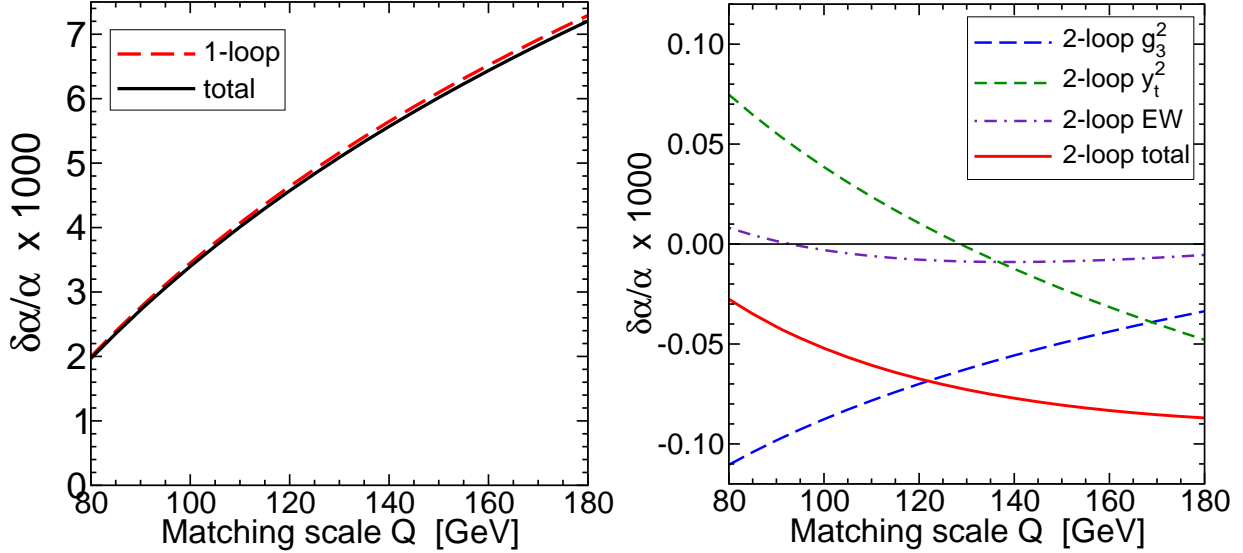

\begin{minipage}[]{0.49\linewidth}
\epsfxsize=\linewidth
\epsffile{thetaEM.eps}
\end{minipage}
\begin{minipage}[]{0.49\linewidth}
\epsfxsize=\linewidth
\epsffile{thetaEM2.eps}
\end{minipage}
\begin{center}
\begin{minipage}[]{0.95\linewidth}
\caption{\label{fig:alphaEMthresh}
Contributions to the matching relation fractional shift in $\alpha$
from decoupling $t,h,Z,W$ in the Standard Model, as a function
of the matching renormalization scale $Q$. 
The left panel shows the dominant 1-loop
contribution (dashed line)  from eq.~(\ref{eq:thetaalpha1}), 
as well as the total (solid line). 
The right panel shows the breakdown of
the total 2-loop contribution from eq.~(\ref{eq:thetaalpha2}) (solid line) into 
the part proportional to $g_3^2$ (long-dashed line), the part
proportional to $y_t^2$ (short-dashed line),  and
the remaining electroweak part (dot-dashed line).}
\end{minipage}
\end{center}
\end{figure}

First, Figure \ref{fig:alphaEMthresh} shows results for
the various contributions to the fractional shift in $\alpha$,
\beq
\delta \alpha/\alpha &\equiv& 
\frac{1}{16 \pi^2} \theta_\alpha^{(1)} + 
\frac{1}{(16 \pi^2)^2} \theta_\alpha^{(2)} + \ldots
.
\eeq
The left panel of Figure \ref{fig:alphaEMthresh} shows the dominant 1-loop
contribution from eq.~(\ref{eq:thetaalpha1}), 
as well as the total from eqs.~(\ref{eq:thetaalpha1}) and (\ref{eq:thetaalpha2}). 
The right panel shows the breakdown of
the 2-loop contribution in eq.~(\ref{eq:thetaalpha2}) into 
the part proportional to $g_3^2$, the part
proportional to $y_t^2$, the remaining pure electroweak part, 
and the total of these 2-loop corrections. 
As might be expected, 
the pure electroweak 2-loop contributions are quite small over the entire
range of $Q$, never exceeding 1 part in $10^5$. The 2-loop $g_3^2$ and $y_t^2$ 
parts are larger, but for lower $Q$
there is significant cancellation between them. 
The total 2-loop contribution ranges from about $-3 \times 10^{-5}$ to
$-9 \times 10^{-5}$, depending on the choice of $Q$. This is comparable to the
present uncertainty on $\Delta \alpha_{\rm had}^{(5)}(m_Z)$ estimated in the 
RPP \cite{Tanabashi:2018oca}, which is $7 \times 10^{-5}$. 
Therefore the total 2-loop correction is just barely numerically
relevant at the present time. If improvements in the
hadronic uncertainty are forthcoming, then the 2-loop corrections will become
correspondingly more significant. However, it seems unlikely that 
further 3-loop corrections to the
matching of $\alpha$ from decoupling $t,h,Z,W$ 
will be needed in the foreseeable future.

\begin{figure}[!t]
\begin{minipage}[]{0.50\linewidth}
\epsfxsize=\linewidth
\epsffile{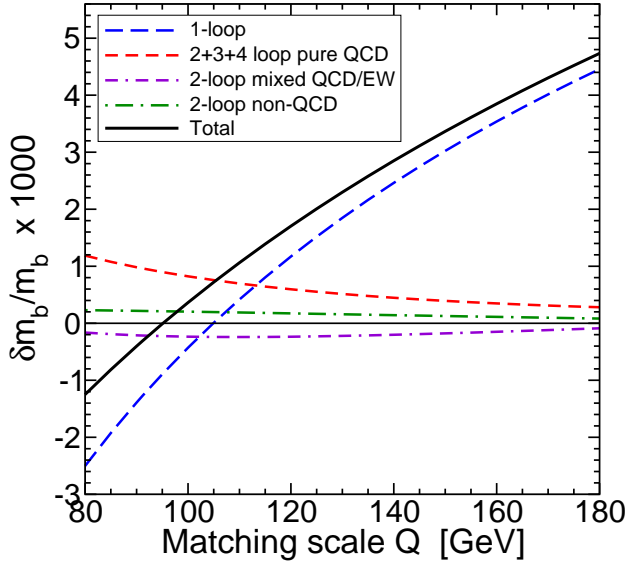}
\end{minipage}
\hspace{0.03\linewidth}
\begin{minipage}[]{0.45\linewidth}
\caption{\label{fig:mbthresh}Contributions to the matching relation fractional
shift in the $\MSbar$ bottom quark mass from decoupling $t,h,Z,W$ in the Standard Model, as a function
of the matching renormalization scale $Q$.
The long-dashed line is the 1-loop contribution
from eq.~(\ref{eq:theta1mb}). 
The short-dashed line is the total QCD (2, 3, and 4-loop) contribution, 
from the $g_3^4$ part of eq.~(\ref{eq:thetam2b}) 
and eqs.~(\ref{eq:thetamq3}) and (\ref{eq:thetamq4}).
The lower and upper dash-dotted lines are from the 
$g_3^2$ (mixed QCD) and $g_3^0$ (non-QCD) parts of eq.~(\ref{eq:thetam2b}), respectively. The solid line is the total.}
\end{minipage}
\end{figure}

\begin{figure}[!t]
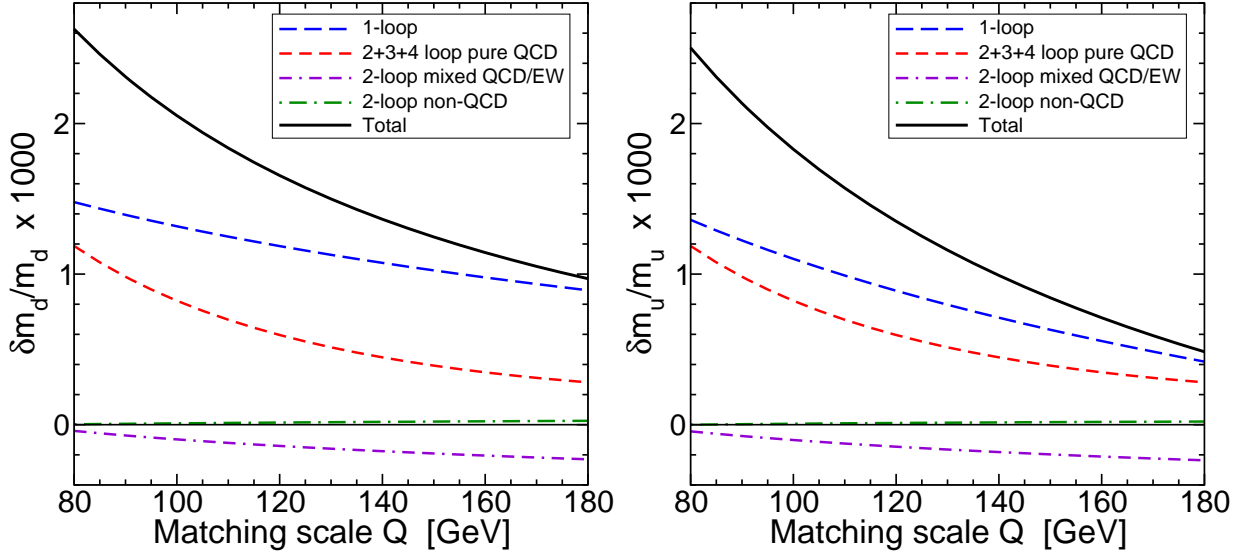

\begin{minipage}[]{0.49\linewidth}
\epsfxsize=\linewidth
\epsffile{thetamd.eps}
\end{minipage}
\begin{minipage}[]{0.49\linewidth}
\epsfxsize=\linewidth
\epsffile{thetamu.eps}
\end{minipage}
\begin{center}
\begin{minipage}[]{0.95\linewidth}
\caption{\label{fig:mudthresh}
Contributions to the matching relation fractional
shift in the $\MSbar$ quark masses 
from decoupling $t,h,Z,W$ in the Standard Model, as a function
of the matching renormalization scale $Q$.
The left panel shows $\delta m_s/m_s = \delta m_d/m_d$, and
the right panel shows $\delta m_c/m_c = \delta m_u/m_u$. In each case,
the long-dashed line is the 1-loop contribution
from eq.~(\ref{eq:theta1mf}). 
The short-dashed line is the total QCD (2, 3, and 4-loop) contribution, 
from the $g_3^4$ part of eq.~(\ref{eq:theta2mf}) 
and eqs.~(\ref{eq:thetamq3}) and (\ref{eq:thetamq4}), and
the lower and upper dash-dotted lines are from the 
$g_3^2$ (mixed QCD) and $g_3^0$ (non-QCD) parts of eq.~(\ref{eq:theta2mf}), respectively. The solid line is the total.}
\end{minipage}
\end{center}
\end{figure}

\begin{figure}[!t]
\begin{minipage}[]{0.49\linewidth}
\epsfxsize=\linewidth
\epsffile{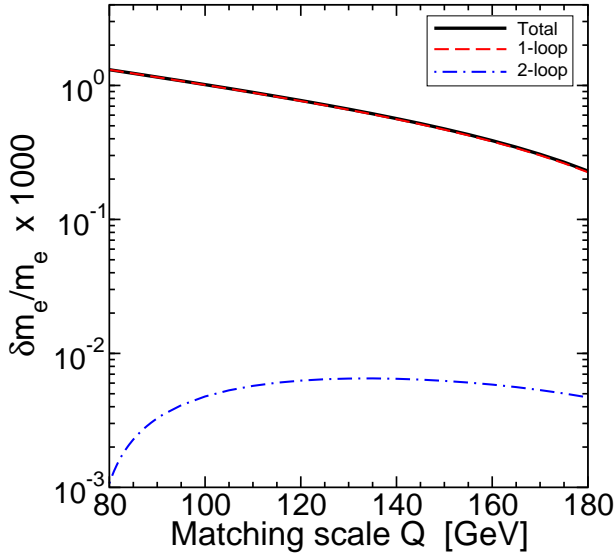}
\end{minipage}
\begin{minipage}[]{0.49\linewidth}
\caption{\label{fig:methresh}
Contributions to the matching relation fractional
shift in the $\MSbar$ charged lepton masses 
from decoupling $t,h,Z,W$ in the Standard Model,
as a function of the matching renormalization scale $Q$.
The solid line is the total, and
the long dashed line hiding just beneath it
is the dominant 1-loop contribution from eq.~(\ref{eq:theta1mf}).
The small difference is the 2-loop contribution from 
eq.~(\ref{eq:theta2mf}), shown as the dot-dashed line.}
\end{minipage}
\end{figure}

The fractional shifts
\beq
\delta m_f/m_f &\equiv&  
\frac{1}{16 \pi^2} \theta_{m_f}^{(1)} + 
\frac{1}{(16 \pi^2)^2} \theta_{m_f}^{(2)} + \ldots
\eeq
are shown in Figures \ref{fig:mbthresh}-\ref{fig:methresh}.
For each of the quark masses, the solid line is the total matching
fractional shift, and the separate contributions 
from 1-loop (to which QCD does not contribute) and the combined 
2, 3, and 4-loop QCD contributions are shown as the long-dashed 
and short-dashed lines, respectively. 
In the case of the bottom quark as shown in Figure \ref {fig:mbthresh}, 
the remaining 2-loop mixed QCD and non-QCD
contributions are each comparable in magnitude to the 3-loop pure QCD part
and much larger than the 4-loop pure QCD part 
(not shown separately), but they have opposite signs from each other
and have a significant cancellation. 
The total fractional shift in $m_b$ from decoupling $t,h,Z,W$ 
is always less than $5 \times 10^{-3}$, and happens 
to be very small for $Q$ near $M_Z$ due to accidental cancellation 
between the different contributions.
(A similar numerical study of the threshold correction for $m_b$
was conducted in ref.~\cite{Bednyakov:2016onn}, 
but with different details because that reference 
uses a different definition of the
high-energy running bottom-quark mass, based on the VEV definition 
$v^2_{\rm on-shell}= 1/\sqrt{2} G_F$.)

In Figure \ref{fig:mudthresh}, 
the results for the down and strange quark masses are shown in the left panel
and for the charm and up quark masses in the right panel.
In both cases, the 2-loop non-QCD corrections are quite tiny, in part
because there is 
no $y_t$ enhancement as there was for the bottom quark. 
The 2-loop mixed QCD corrections are 
larger in magnitude than the 4-loop and comparable to the 3-loop
QCD corrections, but still
less than $2\times 10^{-4}$ over most of the range of choices of $Q$. 
For each of the $c,s,u,d$ quark masses, the total fractional
shifts are slightly larger than
$2 \times 10^{-3}$ for $Q$ near $M_Z$, and decrease with increasing $Q$.
So, they are considerably 
smaller than the present experimental uncertainties in the masses.
This situation is likely to persist for some time, 
pending dramatic improvements in the low-energy 
$\MSbar$ quark mass determinations from e.g.~lattice QCD.

Figure \ref{fig:methresh} shows the results for the 
charged lepton $(\tau, \mu, e)$ masses, for which there are of course 
no QCD-enhanced corrections through 2-loop order.
As expected, the matching is dominated by the 1-loop part, which
contributes of order $2 \times 10^{-4}$ to $2\times 10^{-3}$
to $\delta{m_e}/m_e = \delta{m_\mu}/m_\mu = \delta{m_\tau}/m_\tau$, 
depending on the choice of matching scale $Q$.
The 2-loop contribution to the fractional matching shift 
is seen to be always less than $6 \times 10^{-6}$.
This can be compared to the fractional experimental
uncertainty in the physical
masses of the charged leptons from ref.~\cite{Tanabashi:2018oca}. 
For the tau lepton, this is presently about
$7 \times 10^{-5}$, showing that the 2-loop contribution is 
already safely smaller than the accuracy needed under the most optimistic
of circumstances. For the muon, the fractional uncertainty in the physical
mass is about $2 \times 10^{-8}$ and for the electron 
about $6 \times 10^{-9}$, so in those cases the 2-loop 
(and perhaps even higher loop) threshold matching contributions 
are worthwhile, at least in principle. However, this does not yet take into 
account more subtle parametric uncertainties that are beyond the 
scope of this paper, for example the 
low-energy non-perturbative hadronic
contribution to their pole masses induced through photon
self-energy corrections,
and even small loop effects from $G_F$-suppressed 4-fermion couplings in the
low-energy effective field theory.



\section{Outlook\label{sec:Outlook}}
\setcounter{equation}{0}
\setcounter{figure}{0}
\setcounter{table}{0}
\setcounter{footnote}{1}

In this paper, I have discussed the matching relations for the
renormalizable couplings in the low-energy effective 
$SU(3)_c \times U(1)_{\rm EM}$ gauge theory with 5 quarks and 3 leptons,
when the top quark, Higgs scalar, and $Z$ and $W$ vector bosons are
decoupled together at an $\MSbar$ renormalization scale $Q$. This
simultaneous decoupling ensures that the low-energy effective field
theory has marginal and relevant couplings as part of a consistent
renormalizable gauge theory. Also present in the low-energy theory are
non-renormalizable couplings including 4-fermion terms
for the effective weak interactions; the matching relations for those 
are not discussed in the present paper. The matching relations provide
a connection to the far ultraviolet, fundamental, and complete
definition of the Standard Model. The new results for the matching of
the electromagnetic coupling $\alpha$ and the light quark and lepton
masses augment previously known results for the strong coupling and the
bottom quark mass, and the latter is given here in the tadpole-free
scheme for the VEV, as part of a larger program 
\cite{Martin:2017lqn,Martin:2015eia,Martin:2014bca,Martin:2014cxa,
Martin:2015lxa,Martin:2015rea,Martin:2016xsp}
to relate Standard Model
observables to the underlying Lagrangian parameters in that scheme. The
matching corrections found here are reassuringly small, and in some
cases much smaller than the present experimental uncertainties in the
corresponding observables. They nevertheless are at least useful in
providing informed bounds on the possible sources of theoretical error.
They could become considerably more significant in the future when
experimental uncertainties on input parameters, notably the low-energy
quark masses and non-perturbative contributions to the fine-structure
constant, are reduced.

\vspace{0.35cm}

{\it Acknowledgments:} I am grateful to Jens Erler and Ayres Freitas
for helpful communications regarding ref.~\cite{Tanabashi:2018oca}.
This work was supported in part by the National
Science Foundation grant number PHY-1719273.



\begin{thebibliography}{90}
\baselineskip=13.1pt

\bibitem{Bardeen:1978yd} 
  W.~A.~Bardeen, A.~J.~Buras, D.~W.~Duke and T.~Muta,
  ``Deep Inelastic Scattering Beyond the Leading Order in Asymptotically Free Gauge Theories,''
  Phys.\ Rev.\ D {\bf 18}, 3998 (1978).

\bibitem{Braaten:1981dv} 
  E.~Braaten and J.~P.~Leveille,
  ``Minimal Subtraction and Momentum Subtraction in {QCD} at Two Loop Order,''
  Phys.\ Rev.\ D {\bf 24}, 1369 (1981).

\bibitem{Bollini:1972bi} 
  C.~G.~Bollini and J.~J.~Giambiagi,
  ``Lowest order divergent graphs in nu-dimensional space,''
  Phys.\ Lett.\  {\bf 40B}, 566 (1972).

\bibitem{Ashmore:1972uj} 
  J.~F.~Ashmore,
  ``A Method of Gauge Invariant Regularization,''
  Lett.\ Nuovo Cim.\  {\bf 4}, 289 (1972).

\bibitem{Cicuta:1972jf} 
  G.~M.~Cicuta and E.~Montaldi,
  ``Analytic renormalization via continuous space dimension,''
  Lett.\ Nuovo Cim.\  {\bf 4}, 329 (1972).

\bibitem{tHooft:1972tcz} 
  G.~'t Hooft and M.~J.~G.~Veltman,
  ``Regularization and Renormalization of Gauge Fields,''
  Nucl.\ Phys.\ B {\bf 44}, 189 (1972).

\bibitem{tHooft:1973mfk} 
  G.~'t Hooft,
  ``Dimensional regularization and the renormalization group,''
  Nucl.\ Phys.\ B {\bf 61}, 455 (1973).

\bibitem{Ford:1992pn} 
  C.~Ford, I.~Jack and D.~R.~T.~Jones,
  ``The Standard model effective potential at two loops,''
  Nucl.\ Phys.\ B {\bf 387}, 373 (1992)
  Erratum: [Nucl.\ Phys.\ B {\bf 504}, 551 (1997)]
  [hep-ph/0111190].
``The Effective potential and the differential equations 
method for Feynman integrals,''
  Phys.\ Lett.\ B {\bf 274}, 409 (1992)
  [Erratum-ibid.\ B {\bf 285}, 399 (1992)].

\bibitem{Martin:2001vx} 
  S.~P.~Martin,
  ``Two loop effective potential for a general renormalizable theory and softly broken supersymmetry,''
  Phys.\ Rev.\ D {\bf 65}, 116003 (2002)
  [hep-ph/0111209].

\bibitem{Martin:2013gka} 
  S.~P.~Martin,
  ``Three-loop Standard Model effective potential at leading order in strong and top Yukawa couplings,''
  Phys.\ Rev.\ D {\bf 89}, no. 1, 013003 (2014)
  [arXiv:1310.7553 [hep-ph]].

\bibitem{Martin:2017lqn} 
  S.~P.~Martin,
  ``Effective potential at three loops,''
  Phys.\ Rev.\ D {\bf 96}, no. 9, 096005 (2017)
  [arXiv:1709.02397 [hep-ph]].

\bibitem{Martin:2015eia} 
  S.~P.~Martin,
  ``Four-Loop Standard Model Effective Potential at Leading Order in QCD,''
  Phys.\ Rev.\ D {\bf 92}, no. 5, 054029 (2015)
  [arXiv:1508.00912 [hep-ph]].

\bibitem{Martin:2014bca} 
  S.~P.~Martin,
  ``Taming the Goldstone contributions to the effective potential,''
  Phys.\ Rev.\ D {\bf 90}, no. 1, 016013 (2014)
  [arXiv:1406.2355 [hep-ph]].

\bibitem{Elias-Miro:2014pca} 
  J.~Elias-Miro, J.~R.~Espinosa and T.~Konstandin,
  ``Taming Infrared Divergences in the Effective Potential,''
  JHEP {\bf 1408}, 034 (2014)
  [arXiv:1406.2652 [hep-ph]].

\bibitem{Martin:2016bgz} 
  S.~P.~Martin and D.~G.~Robertson,
  ``Evaluation of the general 3-loop vacuum Feynman integral,''
  Phys.\ Rev.\ D {\bf 95}, no. 1, 016008 (2017)
  [arXiv:1610.07720 [hep-ph]].
The software package {\tt 3VIL} (3-loop Vacuum Integral Library) 
implementing these results is available from 
\begin{verbatim}http://www.niu.edu/spmartin/3VIL/\end{verbatim}

\bibitem{Freitas:2016zmy} 
  A.~Freitas,
  ``Three-loop vacuum integrals with arbitrary masses,''
  JHEP {\bf 1611}, 145 (2016)
  [arXiv:1609.09159 [hep-ph]].
  S.~Bauberger and A.~Freitas,
  ``TVID: Three-loop Vacuum Integrals from Dispersion relations,''
  arXiv:1702.02996 [hep-ph].
  
\bibitem{Fleischer:1980ub} See, for example,
  J.~Fleischer and F.~Jegerlehner,
  ``Radiative Corrections to Higgs Decays in the Extended Weinberg-Salam Model,''
  Phys.\ Rev.\ D {\bf 23}, 2001 (1981).

\bibitem{Martin:2018emo} 
  S.~P.~Martin and H.~H.~Patel,
  ``Two-loop effective potential for generalized gauge fixing,''
  Phys.\ Rev.\ D {\bf 98}, no. 7, 076008 (2018)
  [arXiv:1808.07615 [hep-ph]].
  
\bibitem{Bezrukov:2012sa} 
  F.~Bezrukov, M.~Y.~Kalmykov, B.~A.~Kniehl and M.~Shaposhnikov,
  ``Higgs Boson Mass and New Physics,''
  JHEP {\bf 1210}, 140 (2012)
  [arXiv:1205.2893 [hep-ph]].

\bibitem{Degrassi:2012ry} 
  G.~Degrassi, S.~Di Vita, J.~Elias-Miro, J.~R.~Espinosa, G.~F.~Giudice, G.~Isidori and A.~Strumia,
  ``Higgs mass and vacuum stability in the Standard Model at NNLO,''
  JHEP {\bf 1208}, 098 (2012)
  [arXiv:1205.6497 [hep-ph]].
  
\bibitem{Buttazzo:2013uya} 
  D.~Buttazzo, G.~Degrassi, P.~P.~Giardino, G.~F.~Giudice, F.~Sala, A.~Salvio and A.~Strumia,
  ``Investigating the near-criticality of the Higgs boson,''
  JHEP {\bf 1312}, 089 (2013)
  [arXiv:1307.3536 [hep-ph]].

\bibitem{Martin:2014cxa} 
  S.~P.~Martin and D.~G.~Robertson,
  ``Higgs boson mass in the Standard Model at two-loop order and beyond,''
  Phys.\ Rev.\ D {\bf 90}, no. 7, 073010 (2014)
  [arXiv:1407.4336 [hep-ph]].

\bibitem{Kniehl:2015nwa} 
  B.~A.~Kniehl, A.~F.~Pikelner and O.~L.~Veretin,
  ``Two-loop electroweak threshold corrections in the Standard Model,''
  Nucl.\ Phys.\ B {\bf 896}, 19 (2015)
  [arXiv:1503.02138 [hep-ph]].

\bibitem{Kniehl:2016enc} 
  B.~A.~Kniehl, A.~F.~Pikelner and O.~L.~Veretin,
  ``mr: a C++ library for the matching and running of the Standard Model parameters,''
  Comput.\ Phys.\ Commun.\  {\bf 206}, 84 (2016)
  [arXiv:1601.08143 [hep-ph]].

\bibitem{Sirlin:1980nh} 
  A.~Sirlin,
  ``Radiative Corrections in the $SU(2)_L \times U(1)$ Theory: A Simple Renormalization Framework,''
  Phys.\ Rev.\ D {\bf 22}, 971 (1980).

\bibitem{Marciano:1980pb} 
  W.~J.~Marciano and A.~Sirlin,
  ``Radiative Corrections to Neutrino Induced Neutral Current Phenomena in the $SU(2)_L \times U(1)$ Theory,''
  Phys.\ Rev.\ D {\bf 22}, 2695 (1980)
  Erratum: [Phys.\ Rev.\ D {\bf 31}, 213 (1985)].

\bibitem{Marciano:1983wwa} 
  W.~J.~Marciano and A.~Sirlin,
  ``Testing the Standard Model by Precise Determinations of $W^\pm$ and $Z$ Masses,''
  Phys.\ Rev.\ D {\bf 29}, 945 (1984)
  Erratum: [Phys.\ Rev.\ D {\bf 31}, 213 (1985)].
 
\bibitem{Sirlin:1983ys} 
  A.~Sirlin,
  ``On the O($\alpha^2$) Corrections to $\tau_\mu$, $m_W$, $m_Z$ in the 
  $SU(2)_L \times U(1)$ Theory,''
  Phys.\ Rev.\ D {\bf 29}, 89 (1984).
 
\bibitem{Djouadi:1987gn} 
  A.~Djouadi and C.~Verzegnassi,
  ``Virtual Very Heavy Top Effects in LEP / SLC Precision Measurements,''
  Phys.\ Lett.\ B {\bf 195}, 265 (1987).
 
\bibitem{Djouadi:1987di} 
  A.~Djouadi,
  ``O($\alpha \alpha_S$) Vacuum Polarization Functions of the Standard Model Gauge Bosons,''
  Nuovo Cim.\ A {\bf 100}, 357 (1988).

\bibitem{Consoli:1989fg} 
  M.~Consoli, W.~Hollik and F.~Jegerlehner,
  ``The Effect of the Top Quark on the $M_W$-$M_Z$ Interdependence and Possible Decoupling of Heavy Fermions from Low-Energy Physics,''
  Phys.\ Lett.\ B {\bf 227}, 167 (1989).

\bibitem{Kniehl:1989yc} 
  B.~A.~Kniehl,
  ``Two Loop Corrections to the Vacuum Polarizations in Perturbative QCD,''
  Nucl.\ Phys.\ B {\bf 347}, 86 (1990).

\bibitem{Halzen:1990je} 
  F.~Halzen and B.~A.~Kniehl,
  ``$\Delta r$ beyond one loop,''
  Nucl.\ Phys.\ B {\bf 353}, 567 (1991).

\bibitem{Djouadi:1993ss} 
  A.~Djouadi and P.~Gambino,
  ``Electroweak gauge bosons selfenergies: Complete QCD corrections,''
  Phys.\ Rev.\ D {\bf 49}, 3499 (1994)
  Erratum: [Phys.\ Rev.\ D {\bf 53}, 4111 (1996)]
  [hep-ph/9309298].

\bibitem{Avdeev:1994db} 
  L.~Avdeev, J.~Fleischer, S.~Mikhailov and O.~Tarasov,
  ``${\cal O}(\alpha \alpha_S^2)$ correction to the electroweak $\rho$ parameter,''
  Phys.\ Lett.\ B {\bf 336}, 560 (1994)
  Erratum: [Phys.\ Lett.\ B {\bf 349}, 597 (1995)]
  [hep-ph/9406363].

\bibitem{Chetyrkin:1995ix} 
  K.~G.~Chetyrkin, J.~H.~K\"uhn and M.~Steinhauser,
  ``Corrections of order ${\cal O}(G_F M_t^2 \alpha_S^2)$ to the $\rho$ parameter,''
  Phys.\ Lett.\ B {\bf 351}, 331 (1995)
  [hep-ph/9502291].

\bibitem{Chetyrkin:1995js} 
  K.~G.~Chetyrkin, J.~H.~K\"uhn and M.~Steinhauser,
  ``QCD corrections from top quark to relations between electroweak parameters to order $\alpha_S^2$,''
  Phys.\ Rev.\ Lett.\  {\bf 75}, 3394 (1995)
  [hep-ph/9504413].

\bibitem{Degrassi:1996mg} 
  G.~Degrassi, P.~Gambino and A.~Vicini,
  ``Two loop heavy top effects on the $m_Z - m_W$ interdependence,''
  Phys.\ Lett.\ B {\bf 383}, 219 (1996)
  [hep-ph/9603374].

\bibitem{Degrassi:1996ps} 
  G.~Degrassi, P.~Gambino and A.~Sirlin,
  ``Precise calculation of $M_W$, $\sin^2 \theta_W (M_Z)$, and 
  $\sin^2 \theta_{\rm eff}$(lept),''
  Phys.\ Lett.\ B {\bf 394}, 188 (1997)
  [hep-ph/9611363].

\bibitem{Degrassi:1997iy} 
  G.~Degrassi, P.~Gambino, M.~Passera and A.~Sirlin,
  ``The Role of $M_W$ in precision studies of the standard model,''
  Phys.\ Lett.\ B {\bf 418}, 209 (1998)
  [hep-ph/9708311].

\bibitem{Passera:1998uj} 
  M.~Passera and A.~Sirlin,
  ``Radiative corrections to $W$ and quark propagators in the resonance region,''
  Phys.\ Rev.\ D {\bf 58}, 113010 (1998)
  [hep-ph/9804309].
  
\bibitem{Freitas:2000gg} 
  A.~Freitas, W.~Hollik, W.~Walter and G.~Weiglein,
  ``Complete fermionic two loop results for the $M_W$-$M_Z$ interdependence,''
  Phys.\ Lett.\ B {\bf 495}, 338 (2000)
  Erratum: [Phys.\ Lett.\ B {\bf 570}, no. 3-4, 265 (2003)]
  [hep-ph/0007091].

\bibitem{Freitas:2002ja} 
  A.~Freitas, W.~Hollik, W.~Walter and G.~Weiglein,
  ``Electroweak two loop corrections to the $M_W$-$M_Z$ mass correlation in the standard model,''
  Nucl.\ Phys.\ B {\bf 632}, 189 (2002)
  Erratum: [Nucl.\ Phys.\ B {\bf 666}, 305 (2003)]
  [hep-ph/0202131].

\bibitem{Awramik:2002wn} 
  M.~Awramik and M.~Czakon,
  ``Complete two loop bosonic contributions to the muon lifetime in the standard model,''
  Phys.\ Rev.\ Lett.\  {\bf 89}, 241801 (2002)
  [hep-ph/0208113].
  
\bibitem{Onishchenko:2002ve} 
  A.~Onishchenko and O.~Veretin,
  ``Two loop bosonic electroweak corrections to the muon lifetime and M(Z) - M(W) interdependence,''
  Phys.\ Lett.\ B {\bf 551}, 111 (2003)
  [hep-ph/0209010].

\bibitem{Awramik:2002vu} 
  M.~Awramik, M.~Czakon, A.~Onishchenko and O.~Veretin,
  ``Bosonic corrections to Delta r at the two loop level,''
  Phys.\ Rev.\ D {\bf 68}, 053004 (2003)
  [hep-ph/0209084].
  
\bibitem{Faisst:2003px} 
  M.~Faisst, J.~H.~K\"uhn, T.~Seidensticker and O.~Veretin,
  ``Three loop top quark contributions to the rho parameter,''
  Nucl.\ Phys.\ B {\bf 665}, 649 (2003)
  [hep-ph/0302275].

\bibitem{Awramik:2003ee} 
  M.~Awramik and M.~Czakon,
  ``Complete two loop electroweak contributions to the muon lifetime in the standard model,''
  Phys.\ Lett.\ B {\bf 568}, 48 (2003)
  [hep-ph/0305248].

\bibitem{Awramik:2003rn} 
  M.~Awramik, M.~Czakon, A.~Freitas and G.~Weiglein,
  ``Precise prediction for the W boson mass in the standard model,''
  Phys.\ Rev.\ D {\bf 69}, 053006 (2004)
  [hep-ph/0311148].

\bibitem{Schroder:2005db} 
  Y.~Schroder and M.~Steinhauser,
  ``Four-loop singlet contribution to the rho parameter,''
  Phys.\ Lett.\ B {\bf 622}, 124 (2005)
  [hep-ph/0504055].

\bibitem{Chetyrkin:2006bj} 
  K.~G.~Chetyrkin, M.~Faisst, J.~H.~K\"uhn, P.~Maierhofer and C.~Sturm,
  ``Four-Loop QCD Corrections to the Rho Parameter,''
  Phys.\ Rev.\ Lett.\  {\bf 97}, 102003 (2006)
  [hep-ph/0605201].

\bibitem{Boughezal:2006xk} 
  R.~Boughezal and M.~Czakon,
  ``Single scale tadpoles and O($G_F m_t^2 \alpha_S^3$) corrections to the rho parameter,''
  Nucl.\ Phys.\ B {\bf 755}, 221 (2006)
  [hep-ph/0606232].
            
\bibitem{Jegerlehner:2001fb} 
  F.~Jegerlehner, M.~Y.~Kalmykov and O.~Veretin,
  ``MS versus pole masses of gauge bosons: Electroweak bosonic two loop corrections,''
  Nucl.\ Phys.\ B {\bf 641}, 285 (2002)
  [hep-ph/0105304].

\bibitem{Jegerlehner:2002er} 
  F.~Jegerlehner, M.~Y.~Kalmykov and O.~Veretin,
  ``Full two loop electroweak corrections to the pole masses of gauge bosons,''
  Nucl.\ Phys.\ Proc.\ Suppl.\  {\bf 116}, 382 (2003)
  [hep-ph/0212003].

\bibitem{Jegerlehner:2002em} 
  F.~Jegerlehner, M.~Y.~Kalmykov and O.~Veretin,
  ``$\MSbar$ versus pole masses of gauge bosons. 2. Two loop electroweak fermion corrections,''
  Nucl.\ Phys.\ B {\bf 658}, 49 (2003)
  [hep-ph/0212319].

\bibitem{Degrassi:2014sxa} 
  G.~Degrassi, P.~Gambino and P.~P.~Giardino,
  ``The $m_{\scriptscriptstyle W}-m_{\scriptscriptstyle Z}$ interdependence in the Standard Model: a new scrutiny,''
  JHEP {\bf 1505}, 154 (2015)
  [arXiv:1411.7040 [hep-ph]].

\bibitem{Martin:2015lxa} 
  S.~P.~Martin,
  ``Pole Mass of the W Boson at Two-Loop Order in the Pure $\overline {MS}$ Scheme,''
  Phys.\ Rev.\ D {\bf 91}, no. 11, 114003 (2015)
  [arXiv:1503.03782 [hep-ph]].

\bibitem{Martin:2015rea} 
  S.~P.~Martin,
  ``$Z$-Boson Pole Mass at Two-Loop Order in the Pure $\overline{MS}$ Scheme,''
  Phys.\ Rev.\ D {\bf 92}, no. 1, 014026 (2015)
  [arXiv:1505.04833 [hep-ph]].

\bibitem{Tarrach:1980up} 
  R.~Tarrach,
  ``The Pole Mass in Perturbative QCD,''
  Nucl.\ Phys.\ B {\bf 183}, 384 (1981).

\bibitem{Gray:1990yh} 
  N.~Gray, D.~J.~Broadhurst, W.~Grafe and K.~Schilcher,
  ``Three Loop Relation of Quark $\MSbar$ and Pole Masses,''
  Z.\ Phys.\ C {\bf 48}, 673 (1990).

\bibitem{Chetyrkin:1999ys} 
  K.~G.~Chetyrkin and M.~Steinhauser,
  ``Short distance mass of a heavy quark at order $\alpha_S^3$,''
  Phys.\ Rev.\ Lett.\  {\bf 83}, 4001 (1999)
  [hep-ph/9907509].

\bibitem{Chetyrkin:1999qi} 
  K.~G.~Chetyrkin and M.~Steinhauser,
  ``The Relation between the $\MSbar$ and the on-shell quark mass at order 
  $\alpha_S^3$,''
  Nucl.\ Phys.\ B {\bf 573}, 617 (2000)
  [hep-ph/9911434].

\bibitem{Melnikov:2000qh} 
  K.~Melnikov and T.~v.~Ritbergen,
  ``The Three loop relation between the MS-bar and the pole quark masses,''
  Phys.\ Lett.\ B {\bf 482}, 99 (2000)
  [hep-ph/9912391].

\bibitem{Marquard:2015qpa} 
  P.~Marquard, A.~V.~Smirnov, V.~A.~Smirnov and M.~Steinhauser,
  ``Quark Mass Relations to Four-Loop Order in Perturbative QCD,''
  Phys.\ Rev.\ Lett.\  {\bf 114}, no. 14, 142002 (2015)
  [arXiv:1502.01030 [hep-ph]].

\bibitem{Marquard:2016dcn} 
  P.~Marquard, A.~V.~Smirnov, V.~A.~Smirnov, M.~Steinhauser and D.~Wellmann,
  ``$\overline{\rm MS}$-on-shell quark mass relation up to four loops in QCD and a general SU$(N)$ gauge group,''
  Phys.\ Rev.\ D {\bf 94}, no. 7, 074025 (2016)
  [arXiv:1606.06754 [hep-ph]].

\bibitem{Beneke:1994qe}
  M.~Beneke and V.~M.~Braun,
  ``Naive nonAbelianization and resummation of fermion bubble chains,''
  Phys.\ Lett.\ B {\bf 348}, 513 (1995)
  [hep-ph/9411229].

\bibitem{Ball:1995ni}
  P.~Ball, M.~Beneke and V.~M.~Braun,
  ``Resummation of (beta0 alpha-s)**n corrections in QCD: Techniques and
  applications to the tau hadronic width and the heavy quark pole mass,''
  Nucl.\ Phys.\ B {\bf 452}, 563 (1995)
  [hep-ph/9502300].

\bibitem{Kataev:2015gvt}
  A.~L.~Kataev and V.~S.~Molokoedov,
  ``On the flavour dependence of the $\mathcal{O}(\alpha_s^4)$ correction
  to the relation between running and pole heavy quark masses,''
  Eur.\ Phys.\ J.\ Plus {\bf 131} (2016) no.8,  271
  [arXiv:1511.06898 [hep-ph]].
  ``Multiloop contributions to the $\overline{\rm{MS}}$-on-shell mass 
  relation for heavy quarks in QCD and charged leptons in QED and the
  asymptotic structure of the perturbative QCD series,''
  arXiv:1807.05406 [hep-ph].
  ``Dependence of five and six-loop estimated QCD corrections to the 
  relation between pole and running masses of heavy quarks on the number of
  light flavours,''
  arXiv:1811.02867 [hep-ph].

\bibitem{Beneke:2016cbu}
  M.~Beneke, P.~Marquard, P.~Nason and M.~Steinhauser,
  ``On the ultimate uncertainty of the top quark pole mass,''
  Phys.\ Lett.\ B {\bf 775}, 63 (2017)
  [arXiv:1605.03609 [hep-ph]].

\bibitem{Hoang:2017btd}
  A.~H.~Hoang, C.~Lepenik and M.~Preisser,
  ``On the Light Massive Flavor Dependence of the Large Order Asymptotic
  Behavior and the Ambiguity of the Pole Mass,''
  JHEP {\bf 1709}, 099 (2017)
  [arXiv:1706.08526 [hep-ph]].

\bibitem{Bohm:1986rj} 
  M.~Bohm, H.~Spiesberger and W.~Hollik,
  ``On the One Loop Renormalization of the Electroweak Standard Model and Its Application to Leptonic Processes,''
  Fortsch.\ Phys.\  {\bf 34}, 687 (1986).
  
\bibitem{Hempfling:1994ar} 
  R.~Hempfling and B.~A.~Kniehl,
  ``On the relation between the fermion pole mass and MS Yukawa coupling in the standard model,''
  Phys.\ Rev.\ D {\bf 51}, 1386 (1995)
  [hep-ph/9408313].

\bibitem{Jegerlehner:2003py} 
  F.~Jegerlehner and M.~Y.~Kalmykov,
  ``${\cal O}(\alpha \alpha_S)$ correction to the pole mass of the t quark within the standard model,''
  Nucl.\ Phys.\ B {\bf 676}, 365 (2004)
  [hep-ph/0308216].

\bibitem{Jegerlehner:2003sp} 
  F.~Jegerlehner and M.~Y.~Kalmykov,
  ``${\cal O}(\alpha \alpha_S)$ relation between pole- and MS-bar mass of the t quark,''
  Acta Phys.\ Polon.\ B {\bf 34}, 5335 (2003)
  [hep-ph/0310361].

\bibitem{Faisst:2004gn} 
  M.~Faisst, J.~H.~K\"uhn and O.~Veretin,
  ``Pole versus MS mass definitions in the electroweak theory,''
  Phys.\ Lett.\ B {\bf 589}, 35 (2004)
  [hep-ph/0403026].

\bibitem{Eiras:2005yt} 
  D.~Eiras and M.~Steinhauser,
  ``Two-loop ${\cal O}(\alpha \alpha_S)$ corrections to the on-shell fermion propagator in the standard model,''
  JHEP {\bf 0602}, 010 (2006)
  [hep-ph/0512099].

\bibitem{Jegerlehner:2012kn} 
  F.~Jegerlehner, M.~Y.~Kalmykov and B.~A.~Kniehl,
  ``On the difference between the pole and the $\MSbar$ masses of the top quark at the electroweak scale,''
  Phys.\ Lett.\ B {\bf 722}, 123 (2013)
  [arXiv:1212.4319 [hep-ph]].

\bibitem{Martin:2005ch} 
  S.~P.~Martin,
  ``Fermion self-energies and pole masses at two-loop order in a general renormalizable theory with massless gauge bosons,''
  Phys.\ Rev.\ D {\bf 72}, 096008 (2005)
  [hep-ph/0509115].
  
\bibitem{Kniehl:2014yia} 
  B.~A.~Kniehl and O.~L.~Veretin,
  ``Two-loop electroweak threshold corrections to the bottom and top Yukawa couplings,''
  Nucl.\ Phys.\ B {\bf 885}, 459 (2014)
  Erratum: [Nucl.\ Phys.\ B {\bf 894}, 56 (2015)]
  [arXiv:1401.1844 [hep-ph]].

\bibitem{Martin:2016xsp} 
  S.~P.~Martin,
  ``Top-quark pole mass in the tadpole-free $\overline {MS}$ scheme,''
  Phys.\ Rev.\ D {\bf 93}, no. 9, 094017 (2016)
  [arXiv:1604.01134 [hep-ph]].

\bibitem{MVI}
  M.~E.~Machacek and M.~T.~Vaughn,
  ``Two Loop Renormalization Group Equations in a General Quantum Field Theory. 
  1. Wave Function Renormalization,''
  Nucl.\ Phys.\ B {\bf 222}, 83 (1983).

\bibitem{MVII}
  M.~E.~Machacek and M.~T.~Vaughn,
  ``Two Loop Renormalization Group Equations in a General Quantum Field Theory.
  2. Yukawa Couplings,''
  Nucl.\ Phys.\ B {\bf 236}, 221 (1984).

\bibitem{Jack:1984vj}
  I.~Jack and H.~Osborn,
  ``General Background Field Calculations With Fermion Fields,''
  Nucl.\ Phys.\ B {\bf 249}, 472 (1985).

\bibitem{MVIII}
  M.~E.~Machacek and M.~T.~Vaughn,
  ``Two Loop Renormalization Group Equations in a General Quantum Field Theory.
  3. Scalar Quartic Couplings,''
  Nucl.\ Phys.\ B {\bf 249}, 70 (1985).

\bibitem{Luo:2002ey} 
  M.~x.~Luo and Y.~Xiao,
  ``Two loop renormalization group equations in the standard model,''
  Phys.\ Rev.\ Lett.\  {\bf 90}, 011601 (2003)
  [hep-ph/0207271].

\bibitem{Mihaila:2012fm} 
  L.~N.~Mihaila, J.~Salomon and M.~Steinhauser,
  ``Gauge Coupling Beta Functions in the Standard Model to Three Loops,''
  Phys.\ Rev.\ Lett.\  {\bf 108}, 151602 (2012)
  [arXiv:1201.5868 [hep-ph]].

\bibitem{Chetyrkin:2012rz}
  K.~G.~Chetyrkin and M.~F.~Zoller,
  ``Three-loop $\beta$-functions for top-Yukawa and the Higgs
  self-interaction in the Standard Model,''
  JHEP {\bf 1206}, 033 (2012)
  [1205.2892].

\bibitem{Bednyakov:2012rb} 
  A.~V.~Bednyakov, A.~F.~Pikelner and V.~N.~Velizhanin,
  ``Anomalous dimensions of gauge fields and gauge coupling beta-functions in the Standard Model at three loops,''
  JHEP {\bf 1301}, 017 (2013)
  [arXiv:1210.6873 [hep-ph]].

\bibitem{Bednyakov:2012en} 
  A.~V.~Bednyakov, A.~F.~Pikelner and V.~N.~Velizhanin,
  ``Yukawa coupling beta-functions in the Standard Model at three loops,''
  Phys.\ Lett.\ B {\bf 722}, 336 (2013)
  [arXiv:1212.6829 [hep-ph]].
    
\bibitem{Chetyrkin:2013wya}
  K.~G.~Chetyrkin and M.~F.~Zoller,
  ``$\beta$-function for the Higgs self-interaction in the Standard Model at three-loop level,''
  JHEP {\bf 1304}, 091 (2013)
  [1303.2890].

\bibitem{Bednyakov:2013eba}
  A.~V.~Bednyakov, A.~F.~Pikelner and V.~N.~Velizhanin,
  ``Higgs self-coupling beta-function in the Standard Model at three loops,''
  Nucl.\ Phys.\ B {\bf 875}, 552 (2013)
  [1303.4364].

\bibitem{Bednyakov:2013cpa} 
  A.~V.~Bednyakov, A.~F.~Pikelner and V.~N.~Velizhanin,
  ``Three-loop Higgs self-coupling beta-function in the Standard Model with complex Yukawa matrices,''
  Nucl.\ Phys.\ B {\bf 879}, 256 (2014)
  [arXiv:1310.3806 [hep-ph]].
  
\bibitem{Bednyakov:2014pia} 
  A.~V.~Bednyakov, A.~F.~Pikelner and V.~N.~Velizhanin,
  ``Three-loop SM beta-functions for matrix Yukawa couplings,''
  Phys.\ Lett.\ B {\bf 737}, 129 (2014)
  [arXiv:1406.7171 [hep-ph]].

\bibitem{Chetyrkin:2016ruf} 
  K.~G.~Chetyrkin and M.~F.~Zoller,
  ``Leading QCD-induced four-loop contributions to the $\beta$-function 
  of the Higgs self-coupling in the SM and vacuum stability,''
  JHEP {\bf 1606}, 175 (2016)
  [arXiv:1604.00853 [hep-ph]].

\bibitem{vanRitbergen:1997va} 
  T.~van Ritbergen, J.~A.~M.~Vermaseren and S.~A.~Larin,
  ``The Four loop beta function in quantum chromodynamics,''
  Phys.\ Lett.\ B {\bf 400}, 379 (1997)
  [hep-ph/9701390].

\bibitem{Czakon:2004bu} 
  M.~Czakon,
  ``The Four-loop QCD beta-function and anomalous dimensions,''
  Nucl.\ Phys.\ B {\bf 710}, 485 (2005)
  [hep-ph/0411261].

\bibitem{Baikov:2016tgj} 
  P.~A.~Baikov, K.~G.~Chetyrkin and J.~H.~K\"uhn,
  ``Five-Loop Running of the QCD coupling constant,''
  Phys.\ Rev.\ Lett.\  {\bf 118}, no. 8, 082002 (2017)
  [arXiv:1606.08659 [hep-ph]].

\bibitem{Herzog:2017ohr} 
  F.~Herzog, B.~Ruijl, T.~Ueda, J.~A.~M.~Vermaseren and A.~Vogt,
  ``The five-loop beta function of Yang-Mills theory with fermions,''
  JHEP {\bf 1702}, 090 (2017)
  [arXiv:1701.01404 [hep-ph]].

\bibitem{Tarasov} O.V.~Tarasov, 
``Anomalous Dimensions Of Quark Masses In Three Loop 
Approximation,'' preprint JINR-P2-82-900, (1982), unpublished.

\bibitem{Chetyrkin:1997dh} 
  K.~G.~Chetyrkin,
  ``Quark mass anomalous dimension to ${\cal O}(\alpha_S^4$),''
  Phys.\ Lett.\ B {\bf 404}, 161 (1997)
  [hep-ph/9703278].

\bibitem{Vermaseren:1997fq} 
  J.~A.~M.~Vermaseren, S.~A.~Larin and T.~van Ritbergen,
  ``The four loop quark mass anomalous dimension and the invariant quark mass,''
  Phys.\ Lett.\ B {\bf 405}, 327 (1997)
  [hep-ph/9703284].

\bibitem{Baikov:2014qja} 
  P.~A.~Baikov, K.~G.~Chetyrkin and J.~H.~K\"uhn,
  ``Quark Mass and Field Anomalous Dimensions to ${\cal O}(\alpha_S^5)$,''
  JHEP {\bf 1410}, 076 (2014)
  [arXiv:1402.6611 [hep-ph]].

\bibitem{Weinberg:1980wa} 
  S.~Weinberg,
  ``Effective Gauge Theories,''
  Phys.\ Lett.\  {\bf 91B}, 51 (1980).
B.~A.~Ovrut and H.~J.~Schnitzer,
  ``The Decoupling Theorem and Minimal Subtraction,''
  Phys.\ Lett.\  {\bf 100B}, 403 (1981).
 
\bibitem{Bernreuther:1981sg}  
W.~Bernreuther and W.~Wetzel,
  ``Decoupling of Heavy Quarks in the Minimal Subtraction Scheme,''
  Nucl.\ Phys.\ B {\bf 197}, 228 (1982)
  Erratum: [Nucl.\ Phys.\ B {\bf 513}, 758 (1998)].

\bibitem{Larin:1994va} 
  S.~A.~Larin, T.~van Ritbergen and J.~A.~M.~Vermaseren,
  ``The Large quark mass expansion of $\Gamma (Z^0 \rightarrow \rm{hadrons})$ 
  and $\Gamma (\tau^- \rightarrow \nu_\tau + \rm{hadrons})$ 
  in the order $\alpha_S^3$,''
  Nucl.\ Phys.\ B {\bf 438}, 278 (1995)
  [hep-ph/9411260].

\bibitem{Chetyrkin:1997un} 
  K.~G.~Chetyrkin, B.~A.~Kniehl and M.~Steinhauser,
  ``Decoupling relations to O($\alpha_S^3$) and their connection to low-energy theorems,''
  Nucl.\ Phys.\ B {\bf 510}, 61 (1998)
  [hep-ph/9708255].

\bibitem{Grozin:2011nk} 
  A.~G.~Grozin, M.~Hoeschele, J.~Hoff, M.~Steinhauser,
  ``Simultaneous decoupling of bottom and charm quarks,''
  JHEP {\bf 1109}, 066 (2011)
  [arXiv:1107.5970 [hep-ph]].

\bibitem{Schroder:2005hy} 
  Y.~Schroder and M.~Steinhauser,
  ``Four-loop decoupling relations for the strong coupling,''
  JHEP {\bf 0601}, 051 (2006)
  [hep-ph/0512058].

\bibitem{Chetyrkin:2005ia} 
  K.~G.~Chetyrkin, J.~H.~K\"uhn and C.~Sturm,
  ``QCD decoupling at four loops,''
  Nucl.\ Phys.\ B {\bf 744}, 121 (2006)
  [hep-ph/0512060].

\bibitem{Bednyakov:2014fua} 
  A.~V.~Bednyakov,
  ``On the electroweak contribution to the matching of the strong coupling constant in the SM,''
  Phys.\ Lett.\ B {\bf 741}, 262 (2015)
  [arXiv:1410.7603 [hep-ph]].

\bibitem{Liu:2015fxa} 
  T.~Liu and M.~Steinhauser,
  ``Decoupling of heavy quarks at four loops and effective Higgs-fermion coupling,''
  Phys.\ Lett.\ B {\bf 746}, 330 (2015)
  [arXiv:1502.04719 [hep-ph]].

\bibitem{RunDec} 
  K.~G.~Chetyrkin, J.~H.~K\"uhn and M.~Steinhauser,
  ``RunDec: A Mathematica package for running and decoupling of the strong coupling and quark masses,''
  Comput.\ Phys.\ Commun.\  {\bf 133}, 43 (2000)
  [hep-ph/0004189].
%
  B.~Schmidt and M.~Steinhauser,
  ``CRunDec: a C++ package for running and decoupling of the strong coupling and quark masses,''
  Comput.\ Phys.\ Commun.\  {\bf 183}, 1845 (2012)
  [arXiv:1201.6149 [hep-ph]].
%
  F.~Herren and M.~Steinhauser,
  ``Version 3 of RunDec and CRunDec,''
  Comput.\ Phys.\ Commun.\  {\bf 224}, 333 (2018)
  [arXiv:1703.03751 [hep-ph]].
   
\bibitem{Marciano:1990dp} 
  W.~J.~Marciano and J.~L.~Rosner,
  ``Atomic parity violation as a probe of new physics,''
  Phys.\ Rev.\ Lett.\  {\bf 65}, 2963 (1990)
  Erratum: [Phys.\ Rev.\ Lett.\  {\bf 68}, 898 (1992)].

\bibitem{Degrassi:1990tu} 
  G.~Degrassi, S.~Fanchiotti and A.~Sirlin,
  ``Relations Between the On-shell and $\MSbar$ Frameworks and the $M_W$ - $M_Z$ Interdependence,''
  Nucl.\ Phys.\ B {\bf 351}, 49 (1991).

\bibitem{Fanchiotti:1992tu} 
  S.~Fanchiotti, B.~A.~Kniehl and A.~Sirlin,
  ``Incorporation of QCD effects in basic corrections of the electroweak theory,''
  Phys.\ Rev.\ D {\bf 48}, 307 (1993)
  [hep-ph/9212285].
  
\bibitem{Eidelman:1995ny} 
  S.~Eidelman and F.~Jegerlehner,
  ``Hadronic contributions to $g-2$ of the leptons and to the effective fine structure constant $\alpha (M_Z^2)$,''
  Z.\ Phys.\ C {\bf 67}, 585 (1995)
  [hep-ph/9502298].

\bibitem{Burkhardt:1995tt} 
  H.~Burkhardt and B.~Pietrzyk,
  ``Update of the hadronic contribution to the QED vacuum polarization,''
  Phys.\ Lett.\ B {\bf 356}, 398 (1995).

\bibitem{Martin:1994we} 
  A.~D.~Martin and D.~Zeppenfeld,
  ``A Determination of the QED coupling at the Z pole,''
  Phys.\ Lett.\ B {\bf 345}, 558 (1995)
  [hep-ph/9411377].

\bibitem{Alemany:1997tn} 
  R.~Alemany, M.~Davier and A.~Hocker,
  ``Improved determination of the hadronic contribution to the muon $g-2$ and to $\alpha (M_Z)$ 
  using new data from hadronic tau decays,''
  Eur.\ Phys.\ J.\ C {\bf 2}, 123 (1998)
  [hep-ph/9703220].

\bibitem{Davier:1997vd} 
  M.~Davier and A.~Hocker,
  ``Improved determination of $\alpha (M_Z^2)$ and the anomalous magnetic moment of the muon,''
  Phys.\ Lett.\ B {\bf 419}, 419 (1998)
  [hep-ph/9801361].

\bibitem{Kuhn:1998ze} 
  J.~H.~K\"uhn and M.~Steinhauser,
  ``A Theory driven analysis of the effective QED coupling at $M_Z$,''
  Phys.\ Lett.\ B {\bf 437}, 425 (1998)
  [hep-ph/9802241].
  
\bibitem{Steinhauser:1998rq} 
  M.~Steinhauser,
  ``Leptonic contribution to the effective electromagnetic coupling constant up to three loops,''
  Phys.\ Lett.\ B {\bf 429}, 158 (1998)
  [hep-ph/9803313].
  
\bibitem{Erler:1998sy} 
  J.~Erler,
  ``Calculation of the QED coupling $\alpha (M_Z)$ in the modified minimal subtraction scheme,''
  Phys.\ Rev.\ D {\bf 59}, 054008 (1999)
  [hep-ph/9803453].
  
\bibitem{Erler:1999ug} 
  J.~Erler,
  ``Global fits to electroweak data using GAPP,''
  hep-ph/0005084.
  
\bibitem{Degrassi:2003rw} 
  G.~Degrassi and A.~Vicini,
  ``Two loop renormalization of the electric charge in the standard model,''
  Phys.\ Rev.\ D {\bf 69}, 073007 (2004)
  [hep-ph/0307122].

\bibitem{Tanabashi:2018oca} 
  M.~Tanabashi {\it et al.} [Particle Data Group],
  ``Review of Particle Physics,''
  Phys.\ Rev.\ D {\bf 98}, no. 3, 030001 (2018).

\bibitem{Davier:2017zfy} 
  M.~Davier, A.~Hoecker, B.~Malaescu and Z.~Zhang,
  ``Reevaluation of the hadronic vacuum polarisation contributions to the Standard Model predictions of the muon $g-2$ and ${\alpha (m_Z^2)}$ using newest hadronic cross-section data,''
  Eur.\ Phys.\ J.\ C {\bf 77}, no. 12, 827 (2017)
  [arXiv:1706.09436 [hep-ph]].

\bibitem{Jegerlehner:2017zsb} 
  F.~Jegerlehner,
  ``Variations on Photon Vacuum Polarization,''
  arXiv:1711.06089 [hep-ph].
  
\bibitem{Keshavarzi:2018mgv} 
  A.~Keshavarzi, D.~Nomura and T.~Teubner,
  ``Muon $g-2$ and $\alpha(M_Z^2)$: a new data-based analysis,''
  Phys.\ Rev.\ D {\bf 97}, no. 11, 114025 (2018)
  [arXiv:1802.02995 [hep-ph]].

\bibitem{Kniehl:2004hfa} 
  B.~A.~Kniehl, J.~H.~Piclum and M.~Steinhauser,
  ``Relation between bottom-quark MS-bar Yukawa coupling and pole mass,''
  Nucl.\ Phys.\ B {\bf 695}, 199 (2004)
  [hep-ph/0406254].

\bibitem{Bednyakov:2016onn} 
  A.~V.~Bednyakov, B.~A.~Kniehl, A.~F.~Pikelner and O.~L.~Veretin,
  ``On the $b$-quark running mass in QCD and the SM,''
  Nucl.\ Phys.\ B {\bf 916}, 463 (2017)
  [arXiv:1612.00660 [hep-ph]].
  
\bibitem{Davydychev:1992mt}
  A.~I.~Davydychev and J.~B.~Tausk,
  ``Two loop selfenergy diagrams with different masses and the momentum expansion,''
  Nucl.\ Phys.\ B {\bf 397}, 123 (1993).
A.~I.~Davydychev, V.~A.~Smirnov and J.~B.~Tausk,
  ``Large momentum expansion of two loop selfenergy diagrams 
  with arbitrary masses,''
  Nucl.\ Phys.\ B {\bf 410}, 325 (1993)
  [hep-ph/9307371].
F.~A.~Berends and J.~B.~Tausk,
  ``On the numerical evaluation of scalar two loop selfenergy diagrams,''
  Nucl.\ Phys.\ B {\bf 421}, 456 (1994).

\bibitem{Kniehl:2006bg} 
  B.~A.~Kniehl, A.~V.~Kotikov, A.~I.~Onishchenko and O.~L.~Veretin,
  ``Strong-coupling constant with flavor thresholds at five loops in the $\MSbar$ scheme,''
  Phys.\ Rev.\ Lett.\  {\bf 97}, 042001 (2006)
  [hep-ph/0607202].

\bibitem{Bekavac:2007tk} 
  S.~Bekavac, A.~Grozin, D.~Seidel and M.~Steinhauser,
  ``Light quark mass effects in the on-shell renormalization constants,''
  JHEP {\bf 0710}, 006 (2007)
  [arXiv:0708.1729 [hep-ph]].
  
\bibitem{Bekavac:2009gz}  
  S.~Bekavac, A.~G.~Grozin, D.~Seidel and V.~A.~Smirnov,
  ``Three-loop on-shell Feynman integrals with two masses,''
  Nucl.\ Phys.\ B {\bf 819}, 183 (2009)
  [arXiv:0903.4760 [hep-ph]].

\bibitem{Martin:2006ub} 
  S.~P.~Martin,
  ``Refined gluino and squark pole masses beyond leading order,''
  Phys.\ Rev.\ D {\bf 74}, 075009 (2006)
  [hep-ph/0608026].
      
\bibitem{Baikov:2012zm} 
  P.~A.~Baikov, K.~G.~Chetyrkin, J.~H.~K\"uhn and J.~Rittinger,
  ``Vector Correlator in Massless QCD at Order O($\alpha_S^4$) and the QED beta-function at Five Loop,''
  JHEP {\bf 1207}, 017 (2012)
  [arXiv:1206.1284 [hep-ph]].


\end{thebibliography}
\end{document}